\documentclass{article}

\usepackage{microtype}
\usepackage{graphicx}
\usepackage{subcaption}
\usepackage{booktabs} 

\usepackage{hyperref}

\usepackage[preprint]{modalonet}

\usepackage{amsmath}
\usepackage{amssymb}
\usepackage{mathtools}
\usepackage{amsthm}
\usepackage{algorithm}
\usepackage{algorithmic}
\usepackage{orcidlink}

\usepackage[capitalize,noabbrev]{cleveref}

\usepackage{stfloats}     

\usepackage{tikz}
\usepackage{pgfplots}
\pgfplotsset{compat=1.17}
\usetikzlibrary{arrows.meta,backgrounds,fit,calc,positioning,patterns,decorations.pathreplacing}
\definecolor{ioblue}{HTML}{AFCDE9}\definecolor{oporange}{HTML}{F4A85C}
\definecolor{cream}{HTML}{F8EAA6}\definecolor{tranyellow}{HTML}{F1D67E}
\definecolor{summred}{HTML}{E7693B}\definecolor{polenode}{HTML}{A9C0DF}
\definecolor{edgecol}{HTML}{33414E}\definecolor{inkcol}{HTML}{1B1B2A}
\definecolor{beamcol}{HTML}{4A5A6A}\definecolor{platefill}{HTML}{DCE6F2}
\definecolor{dimcol}{HTML}{555555}
\definecolor{trainfill}{HTML}{E8F0FE}\definecolor{trainline}{HTML}{4285F4}
\definecolor{lossfill}{HTML}{FEF7E0}\definecolor{lossline}{HTML}{F9AB00}
\definecolor{extrfill}{HTML}{F3E8FD}\definecolor{extrline}{HTML}{A142F4}
\definecolor{outfill}{HTML}{E6F4EA}\definecolor{outline}{HTML}{34A853}
\definecolor{datafill}{HTML}{E3EEFC}\definecolor{dataline}{HTML}{1A73E8}
\definecolor{sgfill}{HTML}{F8F9FB}\definecolor{sgline}{HTML}{C8CDD6}
\definecolor{edgec}{HTML}{5F6671}\definecolor{ink}{HTML}{202437}
\definecolor{nodeln}{HTML}{55606E}\definecolor{trnbg}{HTML}{F1F5FA}\definecolor{trnbd}{HTML}{C7D4E3}\definecolor{extbg}{HTML}{F7F4F0}\definecolor{extbd}{HTML}{DACEBE}\definecolor{lossbd}{HTML}{C49A4E}\definecolor{lossbg}{HTML}{FBF4E2}\definecolor{accfill}{HTML}{2E6E5A}\definecolor{poleX}{HTML}{B5232E}\definecolor{databd}{HTML}{6E96C8}
\pgfplotsset{
  colormap={rdbu}{rgb255=(5,48,97) rgb255=(33,102,172) rgb255=(67,147,195)
                  rgb255=(146,197,222) rgb255=(209,229,240) rgb255=(247,247,247)
                  rgb255=(253,219,199) rgb255=(244,165,130) rgb255=(214,96,77)
                  rgb255=(178,24,43) rgb255=(103,0,31)},
  modeaxis/.style={hide axis,scale only axis,width=1.2cm,height=1.2cm,
                   view={0}{90},enlargelimits=false,
                   point meta min=-1,point meta max=1,colormap name=rdbu},
}

\theoremstyle{plain}

\theoremstyle{definition}

\theoremstyle{remark}

\newcommand{\bs}{\mathbf{s}}
\newcommand{\xhat}{\hat{x}}

\motitlerunning{Learning Structural Eigenmodes with ModalONet}

\begin{document}

\twocolumn[
  \motitle{Learning Structural Eigenmodes with Modal Operator Network (ModalONet)}

  \mosetsymbol{equal}{*}

  \begin{moauthorlist}
    \moauthor{Saad Waheed \orcidlink{0009-0002-8700-6140}}{DSSL}
    \moauthor{Shabbir Ahmed \orcidlink{0000-0001-8296-6025}}{DSSL}
  \end{moauthorlist}

  \moaffiliation{DSSL}{Dynamical Systems and Signals Lab (DSSL), Department of Mechanical Engineering,
    South Dakota State University, Brookings, SD, USA}

  \mocorrespondingauthor{Shabbir Ahmed}{shabbir.ahmed@sdstate.edu}

  \mokeywords{Neural Operators, DeepONet, Laplace Neural Operator, Modal Analysis, Structural Dynamics, Mode Shapes}

  \vskip 0.3in
]

\printAffiliationsAndNotice{}

\begin{abstract}
  Neural operators such as the deep operator network (DeepONet) and the Laplace neural operator (LNO) are effective surrogates, but they are almost exclusively trained to reproduce the \emph{forward response} of a system rather than its intrinsic structure, such as the modal properties of a structural system. We introduce ModalONet, which puts operator learning to a different use such as recovering the \emph{modal basis} namely: mode shapes, natural frequencies, and damping ratios of a dynamical system directly from its response field, with no eigensolver and no labeled modes. Our key observation is that the DeepONet branch--trunk factorization is itself a learnable form of modal superposition: the trunk supplies continuous, mesh-free mode shapes, while an LNO branch supplies the modal coordinates in pole--residue form, so that each learned pole yields a natural frequency and a damping ratio. Training uses the response field alone, under a composite loss of reconstruction, orthonormality, and temporal  projection consistency. The degenerate (equal-frequency) modes are resolved by a separable trunk and a shared frequency parameter, with post-hoc log-envelope regression improves the damping ratio estimates. Across four structural systems, namely: simply supported and cantilever Euler--Bernoulli beams and rectangular and (degenerate) square Kirchhoff plates, the ModalONet recovers the analytical modal basis with modal assurance criterion (MAC) values of at least $\mathbf{0.998}$ for every mode shape, natural frequency errors within $5\%$, and damping ratio errors within $7\%$, demonstrating the potential of neural operators as accurate and interpretable tools for modal identification.
   
\end{abstract}

\section{Introduction}
\label{sec:intro}

The dynamic behavior of an engineered structure such as a bridge under traffic, an aircraft wing subjected to gust loading, or a turbine blade in operation is most naturally described by its modal parameters: the natural frequencies, damping ratios, and spatial mode shapes that characterize its vibration. For linear, proportionally damped systems, any structural response can be represented as a superposition of the mode shapes, which thus constitute the natural coordinate basis of structural dynamics. Beyond governing resonance and transient response, these parameters serve as calibration targets in finite-element model updating and as damage-sensitive features in structural health monitoring (SHM), where shifts in natural frequencies and distortions of mode shapes are primary indicators of structural degradation \citep{giurgiutiu2014shm,ahmed2025functional,ahmed2025autoregressive,ahmed2024active}. Recovering these modal parameters from measured structural responses is therefore a fundamental inverse problem: the accuracy of downstream tasks such as model updating, damage detection, and response prediction depends directly on how faithfully they are identified.

Classically, these modal parameters are obtained in one of two ways. The first solves the eigenproblem of a discretized finite-element model, which yields natural frequencies and mode shapes, and presupposes accurate geometry, material properties, and boundary conditions, requiring re-calibration whenever any of these change \cite{rizmi2026vision}. The second identifies the modal parameters directly from measured vibration signals, following the methods of experimental or operational modal analysis \citep{ewins2000modal, brincker2001modal, peeters2001stochastic}. A rich family of estimators has been developed for this purpose, spanning time-domain methods such as the eigensystem realization algorithm and stochastic subspace identification \citep{juang1985era, peeters2001stochastic} and frequency-domain techniques such as frequency-domain decomposition \citep{brincker2001modal}. These methods are mature and highly successful across a wide range of engineering structures. However, practical modal identification often involves user-dependent decisions, including model-order selection and the separation of physical modes from numerical or noise-induced modes. Furthermore, experimentally identified mode shapes are naturally obtained as discrete vectors at measured degrees of freedom, requiring additional expansion procedures to obtain full-field representations. These limitations motivate data-driven approaches that learn modal representations directly from measured responses, recover continuous spatial mode shapes, and reduce reliance on manual identification steps.

Data-driven decomposition methods offer a complementary route for extracting coherent spatio-temporal patterns directly from measured responses. Proper orthogonal decomposition (POD) computes the spatial basis via singular-value decomposition that optimally captures the response energy in the mean-square sense \citep{berkooz1993pod}. Dynamic mode decomposition (DMD) extracts spatio-temporal modes associated with the eigenvalues of a best-fit linear evolution operator, yielding characteristic frequencies and growth or decay rates \citep{schmid2010dmd}. These methods have proven highly successful for reduced-order modeling, coherent-structure discovery, and, in the structural setting, damage-sensitive feature extraction. Their correspondence with the modal parameters, however, is conditional: POD modes approximate the structural mode shapes only under restrictive mass and excitation conditions \citep{feeny1998pod}, and DMD recovers them faithfully when the dynamics are linear and observed through displacement-proportional measurements, with accuracy degrading under noise, nonlinearity, or indirect observables such as raw video intensities.

Previously, neural network--based models have been utilized to extract modal information from data. Bao et al. utilized a mechanics-informed neural network model to extract modal information from measured responses. The study processed the time-domain data with the short-time Fourier transform (STFT) to obtain a time-frequency representation, from which single source points (SSPs) were extracted and used as input to the network \citep{bao2024mechanics}. Su et al. utilized a convolutional neural network to extract modal parameters, where uncertainty diagrams of the modal parameters, estimated by stochastic subspace identification, were used as input to the network \citep{su2020automatic}. Jian et al. utilized a graph neural network to identify modal parameters, using the power spectral density of the time-domain data as input \citep{jian2025gnn}. Liu et al. used only time-domain data as input to the neural network and extracted the modal parameters through a suitable choice of loss function \citep{liu2021blind}. Liu et al. used a 1-D convolutional neural network coupled with a long short-term memory (LSTM) network, with time-domain signals obtained from video recordings of a vibrating structure as input \citep{liu2024cvmodal}. Liu and Bao extended the mechanics-informed approach to the identification of closely spaced modes \citep{liu2025closely}, and Lai et al. proposed neural modal ordinary differential equations, which embed modal structure in learned dynamics for high-dimensional monitored structures \citep{lai2022neural}. Sadhu et al. reviewed output-only modal identification based on blind source separation \citep{sadhu2017review}, a family recently accelerated by AI-driven source separation for fast operational modal analysis \citep{hernandez2024blind}. As these methods rely on conventional neural network architectures, they may suffer from the requirement of fixed-length inputs and fixed geometry and boundary conditions. In many practical applications, however, modal parameters must be estimated for arbitrary geometries and boundary conditions, solely from time-domain data.

Neural networks have also been utilized to solve eigenproblems; however, these methods assume that the operator is known. Ben-Shaul et al., Deng et al., and Han et al. developed deep eigensolvers that recover the eigenfunctions of a given differential operator or kernel by minimizing a Rayleigh-quotient or residual objective \citep{benshaul2023deep, deng2022neuralef, han2020solving}. Recently, Yang et al. proposed a neural operator that predicts the Laplace--Beltrami eigenspace of a shape directly from its geometry \citep{yang2026neural}. However, these methods solve the forward eigenproblem of an operator supplied in closed-form, which is the
inverse of the problem of identifying modes from measured response, and they do not account for damping.

In contrast, operator learning has been developed to learn mappings between function spaces rather than fitting a single solution, which supplies the missing notion of transfer. Chen and Chen established a universal approximation theorem for operators \citep{chen1995universal}, based on which Lu et al. proposed the deep operator network (DeepONet), which factorizes an operator into a branch network acting on the input function and a trunk network acting on the query coordinate \citep{lu2021learning}. Li et al. proposed the Fourier neural operator, which parameterizes the operator in the spectral domain \citep{li2021fourier}, and Kovachki et al. developed a general theory of neural operators
\citep{kovachki2023neural}. Raissi et al. introduced physics-informed training, in which the governing equations are embedded directly in the loss function \citep{raissi2019physics}, Wang et al. extended this idea to operator learning \citep{wang2021learning}, and Lu et al. reported fair comparisons across different operator families \citep{lu2022comprehensive}. These operators have been applied to a wide range of problems. Goswami et al. applied a variational DeepONet to fracture mechanics \citep{goswami2022variational}. He et al. trained an operator to predict
the full-field response of structures under time-dependent loads \citep{he2024sequential}, and Cao et al. predicted the motion of floating offshore structures \citep{cao2024floating}. De et al. and Goswami et al. addressed hysteretic and stochastic nonlinear responses \citep{de2024bifidelity, goswami2025neural}, Garg et al. assessed reliability under stochastic loading \citep{garg2022assessment}, Liu et al. learned causal responses of linear systems \citep{liu2024causality}, and Kaewnuratchadasorn et al. predicted bridge response for structural health monitoring \citep{kaewnuratchadasorn2024neural}. However, in all these studies, the operator learns the forward solution, i.e., the mapping from excitation or parameters to
response, and the modal structure, if present at all, remains an implicit by-product
rather than an output of the network.

To close this gap, we propose a neural operator whose output is the modal basis itself, recovered from response data without any eigensolver and without labeled modes. The key observation is structural: modal superposition, i.e., a finite sum of products of a temporal factor and a spatial factor, is exactly the form of the DeepONet branch--trunk output, and therefore a DeepONet trained to reconstruct a response field is already performing a modal decomposition. Two ingredients render its latent factors physical: spatial orthonormality of the trunk, which turns the spatial factors into continuous, mesh-free mode shapes, and a Laplace neural operator (LNO) branch \citep{cao2024laplace}, whose pole--residue parameterization
represents a damped modal coordinate exactly. Reading each learned pole as $-\sigma_i + j\omega_{d,i}$ yields the natural frequency and damping ratio directly, so that a single architecture returns all three modal quantities at once. To the best of our knowledge, this is the first neural operator to recover the structural modal basis from response data. The proposed method differs from population-based graph networks, which remain grid-bound and post-process frequency and damping \citep{jian2025modal}, and from neural eigensolvers, which require the operator in closed-form and target geometric, undamped spectra \citep{yang2026neural,
benshaul2023deep}.

\textbf{Contributions.} The main contributions of this work are as follows.\\
(i)~We recast modal identification as an operator learning problem, showing that the DeepONet branch--trunk factorization is a learnable form of modal superposition, and we use it to recover the full modal basis---mode shapes, natural frequencies, and damping ratios---from response data rather than a forward solution.\\
(ii)~We train the network in an unsupervised manner on response fields alone, using an orthonormality constraint and a temporal projection consistency term; we resolve degenerate equal-frequency modes via the separable trunk, a shared (tied) frequency parameter per degenerate group, and an assignment step; and we refine damping ratio post hoc by an SNR-weighted log-envelope regression on the modal projections.\\
(iii)~We validate the method on Euler--Bernoulli beams and Kirchhoff plates, recovering the analytical modal basis at MAC value at least $0.998$, including the degenerate square plate, with every latent dimension interpretable as a physical mode.

\section{Method}
\label{sec:method}

\subsection{Structural modal analysis}
\label{sec:modal}
\label{sec:expansion}

We consider small-amplitude vibration of linear elastic structures, whose motion is governed by a linear partial differential equation in space and time. For a slender beam of length $l$, bending stiffness $EI$, and mass per unit length $m=\rho A$, the transverse displacement $w(x,t)$ obeys the Euler--Bernoulli equation \citep{giurgiutiu2014shm}
\begin{equation}
  EI\,\frac{\partial^4 w}{\partial x^4} + m\,\frac{\partial^2 w}{\partial t^2}
  = f(x,t),
  \label{eq:eb}
\end{equation}
with $f$ a distributed transverse load. For a thin (Kirchhoff) plate of thickness $h$, density $\rho$, and Poisson ratio $\nu$, the transverse displacement obeys
\begin{equation}
  D\,\nabla^4 w + \rho h\,\frac{\partial^2 w}{\partial t^2} = 0,
  \qquad
  D = \frac{E h^3}{12(1-\nu^2)},
  \label{eq:plate}
\end{equation}
with biharmonic operator $\nabla^4=\partial_{xxxx}+2\partial_{xxyy}+\partial_{yyyy}$ and flexural rigidity $D$ \citep{giurgiutiu2014shm}.

\begin{figure*}[tbp]
  \vskip 0.1in
  \centering
    \begin{tikzpicture}[
  >={Stealth[length=2mm,width=1.6mm]},
  beam/.style={beamcol,line width=2.2pt,line cap=round},
  dim/.style={dimcol,line width=.5pt,{Stealth[length=1.6mm]}-{Stealth[length=1.6mm]}},
  ext/.style={dimcol,line width=.35pt},
  bc/.style={inkcol,font=\footnotesize},
  ttl/.style={inkcol,font=\small\bfseries},
]
\tikzset{
  pin/.pic={
    \draw[fill=gray!18,line width=.8pt] (0,0)--(-0.24,-0.42)--(0.24,-0.42)--cycle;
    \draw[line width=.8pt] (-0.40,-0.42)--(0.40,-0.42);
    \foreach \x in {-0.30,-0.15,0,0.15,0.30}
      \draw[line width=.5pt] (\x,-0.42)--(\x-0.13,-0.58);},
  roller/.pic={
    \draw[fill=gray!18,line width=.8pt] (0,0)--(-0.24,-0.30)--(0.24,-0.30)--cycle;
    \draw[fill=white,line width=.6pt] (-0.12,-0.39) circle (0.085);
    \draw[fill=white,line width=.6pt] ( 0.12,-0.39) circle (0.085);
    \draw[line width=.8pt] (-0.40,-0.48)--(0.40,-0.48);
    \foreach \x in {-0.30,-0.15,0,0.15,0.30}
      \draw[line width=.5pt] (\x,-0.48)--(\x-0.13,-0.64);},
  fixed/.pic={
    \fill[pattern=north east lines] (-0.20,-0.55) rectangle (0,0.55);
    \draw[line width=1pt] (0,-0.55)--(0,0.55);},
}

\begin{scope}[shift={(0,0)}]
  \draw[beam] (0,0)--(4,0);
  \pic at (0,0) {pin};
  \pic at (4,0) {roller};
  \draw[ext] (0,-0.62)--(0,-1.05); \draw[ext] (4,-0.66)--(4,-1.05);
  \draw[dim] (0,-1.0)--(4,-1.0);
  \node[bc,fill=white,inner sep=1pt] at (2,-1.0) {$l$};
  \node[bc,anchor=south] at (0,0.18) {$w=0,\ w''=0$};
  \node[bc,anchor=south] at (4,0.18) {$w=0,\ w''=0$};
  \node[ttl] at (2,-1.55) {(a)};
\end{scope}

\begin{scope}[shift={(6.6,0)}]
  \draw[beam] (0,0)--(4,0);
  \pic at (0,0) {fixed};
  \draw[beamcol,line width=.8pt,fill=white] (4,0) circle (0.06);
  \draw[ext] (0,-0.6)--(0,-1.05); \draw[ext] (4,-0.2)--(4,-1.05);
  \draw[dim] (0,-1.0)--(4,-1.0);
  \node[bc,fill=white,inner sep=1pt] at (2,-1.0) {$l$};
  \node[bc,anchor=south east] at (-0.12,0.30) {$w=0$};
  \node[bc,anchor=north east] at (-0.12,-0.05) {$w'=0$};
  \node[bc,anchor=south] at (4,0.18) {$w''=0,\ w'''=0$};
  \node[ttl] at (2,-1.55) {(b)};
\end{scope}

\begin{scope}[shift={(0,-5.1)}]
  \def\aa{3.6}\def\bb{2.2}
  \begin{scope}
    \clip (-0.16,-0.16) rectangle (\aa+0.16,\bb+0.16);
    \fill[pattern=north east lines,pattern color=gray!70]
      (-0.16,-0.16) rectangle (\aa+0.16,\bb+0.16);
  \end{scope}
  \fill[platefill,draw=beamcol,line width=1.4pt] (0,0) rectangle (\aa,\bb);
  \draw[ext] (0,-0.30)--(0,-0.85); \draw[ext] (\aa,-0.30)--(\aa,-0.85);
  \draw[dim] (0,-0.8)--(\aa,-0.8); \node[bc,fill=white,inner sep=1pt] at (\aa/2,-0.8) {$a$};
  \draw[ext] (-0.30,0)--(-0.85,0); \draw[ext] (-0.30,\bb)--(-0.85,\bb);
  \draw[dim] (-0.8,0)--(-0.8,\bb); \node[bc,fill=white,inner sep=1pt] at (-0.8,\bb/2) {$b$};
  \draw[->,line width=.7pt] (0.2,0.2)--(1.0,0.2) node[bc,right=-1pt]{$x$};
  \draw[->,line width=.7pt] (0.2,0.2)--(0.2,1.0) node[bc,above=-1pt]{$y$};
  \node[bc] at (\aa/2,\bb/2) {$w=0$ on $\partial\Omega$};
  \node[ttl] at (\aa/2,-1.45) {(c)};
\end{scope}

\begin{scope}[shift={(6.6,-5.1)}]
  \def\aa{2.2}
  \begin{scope}
    \clip (-0.16,-0.16) rectangle (\aa+0.16,\aa+0.16);
    \fill[pattern=north east lines,pattern color=gray!70]
      (-0.16,-0.16) rectangle (\aa+0.16,\aa+0.16);
  \end{scope}
  \fill[platefill,draw=beamcol,line width=1.4pt] (0,0) rectangle (\aa,\aa);
  \draw[ext] (0,-0.30)--(0,-0.85); \draw[ext] (\aa,-0.30)--(\aa,-0.85);
  \draw[dim] (0,-0.8)--(\aa,-0.8); \node[bc,fill=white,inner sep=1pt] at (\aa/2,-0.8) {$a$};
  \draw[ext] (-0.30,0)--(-0.85,0); \draw[ext] (-0.30,\aa)--(-0.85,\aa);
  \draw[dim] (-0.8,0)--(-0.8,\aa); \node[bc,fill=white,inner sep=1pt] at (-0.8,\aa/2) {$a$};
  \node[bc,align=center] at (\aa/2,\aa/2)
     {$a=b$\\[1pt]$\omega_{(2,1)}=\omega_{(1,2)}$};
  \node[ttl] at (\aa/2,-1.45) {(d)};
\end{scope}
\end{tikzpicture}
    \caption{The four structural configurations studied here are shown: (a) simply supported beam; (b) cantilever beam; (c) rectangular SSSS plate ($a{=}1.5$, $b{=}1.0$); (d) square SSSS plate ($a{=}b{=}1.0$), whose modes $(2,1)$ and $(1,2)$ are degenerate. Hatched symbols denote supports; in-figure annotations state the boundary conditions and dimensions of each case.\\}
    \label{fig:bc}
  \vskip -0.15in
\end{figure*}
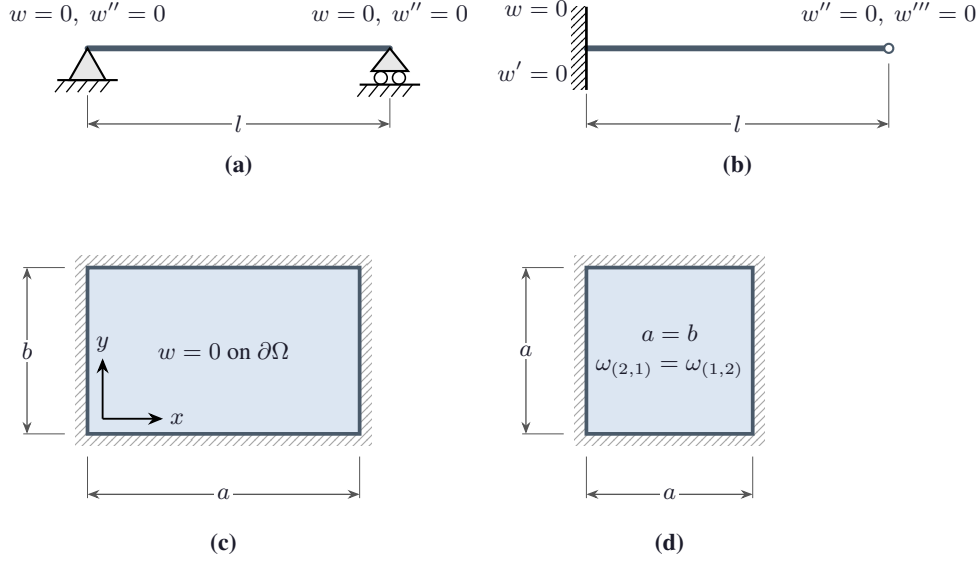

Seeking free, time-harmonic motion $w=\phi(\bs)e^{j\omega t}$ separates space from time and reduces either equation to a spatial eigenproblem whose eigenpairs $(\omega_i,\phi_i)$ are the natural frequencies and mode shapes, selected by the boundary conditions. These eigenfunctions form a complete basis for the response, so any free vibration is a superposition of them,
\begin{equation}
  w(\bs,t) = \sum_{i=1}^{\infty} q_i(t)\,\phi_i(\bs),
  \label{eq:expansion}
\end{equation}
in which each mode shape $\phi_i(\bs)$ is a fixed spatial pattern and each modal coordinate $q_i(t)$ carries its time-varying amplitude. The eigenproblem determines each mode shape only up to an arbitrary scale, and distinct modes are meaningful only if they are mutually independent. We therefore impose the orthonormality condition
\begin{equation}
  \langle \phi_p,\phi_q\rangle
  = \int_\Omega \phi_p(\bs)\,\phi_q(\bs)\,\mathrm{d}\bs = \delta_{pq},
  \label{eq:ortho}
\end{equation}
which holds under the $L^2(\Omega)$ inner product for a structure of uniform mass distribution, where the mass-weighted orthogonality of the modes reduces to the unweighted $L^2$ product \citep{giurgiutiu2014shm}.

Orthonormality decouples the equations of motion, so each modal coordinate is governed by an independent single-degree-of-freedom oscillator. \Cref{eq:eb,eq:plate} are undamped, whereas real structures dissipate energy; under the standard assumption of proportional (classical) damping, the mode shapes of the undamped problem also diagonalize the damping term, so each coordinate acquires its own modal damping ratio $\zeta_i$ while $\phi_i$ and $\omega_i$ are unchanged \citep{ewins2000modal}. The free response of mode $i$ is then the decaying oscillation
\begin{equation}
  \begin{aligned}
  q_i(t) &= A_i\,e^{-\zeta_i\omega_i t}
            \cos\!\left(\omega_{d,i}\,t+\varphi_i\right),\\[2pt]
  \omega_{d,i} &= \omega_i\sqrt{1-\zeta_i^{2}},
  \end{aligned}
  \label{eq:modalcoord}
\end{equation}
with amplitude $A_i$ and phase $\varphi_i$ fixed by the initial conditions and $\omega_{d,i}$ the damped natural frequency. \Cref{eq:expansion,eq:modalcoord} together state the structure we exploit throughout: a response field is a sum of products of a spatial factor and a decaying-oscillatory temporal factor.

The eigenpairs themselves follow from the boundary conditions. For a uniform beam, \cref{eq:eb} gives $\phi''''-\gamma^4\phi=0$ with $\gamma^4=m\omega^2/EI$, solved by a combination of $\sin\gamma x,\cos\gamma x,\sinh\gamma x,\cosh\gamma x$. Three boundary-condition cases cover the four test systems used in this work.

\textbf{Simply supported (pinned) beam.} With $w=\partial_{xx}w=0$ at both ends, the analytical mode shapes and natural frequencies are
\begin{equation}
  \phi_n(x)=\sqrt{\tfrac{2}{l}}\,\sin\frac{n\pi x}{l},
  \quad
  \omega_n=\Big(\frac{n\pi}{l}\Big)^{\!2}\sqrt{\frac{EI}{m}},
  \label{eq:ssbeam}
\end{equation}
for $n=1,2,\dots$ \citep{giurgiutiu2014shm}.

\textbf{Cantilever (clamped--free) beam.} With $w=\partial_x w=0$ at $x=0$ and $\partial_{xx}w=\partial_{xxx}w=0$ at $x=l$, the natural frequencies follow from the eigenvalues $z_n=\gamma_n l$ of the characteristic equation
\begin{equation}
  \cos z\,\cosh z + 1 = 0,
  \quad z_n \approx 1.875,\,4.694,\,7.855,\dots
  \label{eq:canteig}
\end{equation}
as $\omega_n=z_n^2\sqrt{EI/ml^4}$, and the analytical mode shapes are
\begin{equation}
  \begin{aligned}
  \phi_n(x)={}&\tfrac{1}{\sqrt{l}}\big[\cosh\gamma_n x-\cos\gamma_n x\\
  &\quad-\beta_n\!\left(\sinh\gamma_n x-\sin\gamma_n x\right)\big],\\[2pt]
  \beta_n={}&\frac{\cosh\gamma_n l+\cos\gamma_n l}
                  {\sinh\gamma_n l+\sin\gamma_n l},
  \end{aligned}
  \label{eq:cantshape}
\end{equation}
for $n=1,2,\dots$ \citep{giurgiutiu2014shm}. Unlike the sinusoids of \cref{eq:ssbeam}, these shapes combine hyperbolic and trigonometric terms, making them a stricter test of a learned spatial representation.

\textbf{Simply supported rectangular plate (SSSS).} With $w=0$ and zero bending moment on all edges of an $a\times b$ plate, \cref{eq:plate} admits the analytical mode shapes and natural frequencies
\begin{equation}
  \begin{aligned}
  \phi_{mn}(x,y)&=\tfrac{2}{\sqrt{ab}}\sin\tfrac{m\pi x}{a}\,\sin\tfrac{n\pi y}{b},\\[2pt]
  \omega_{mn}&=\pi^2\sqrt{\tfrac{D}{\rho h}}
  \left[\Big(\tfrac{m}{a}\Big)^{2}\!+\Big(\tfrac{n}{b}\Big)^{2}\right]\!,
  \end{aligned}
  \label{eq:plateshape}
\end{equation}
for $m,n=1,2,\dots$ \citep{giurgiutiu2014shm}, with each mode indexed by the number of half-waves $m$ along $x$ and $n$ along $y$.

A \emph{degenerate} case arises when the plate is square ($a=b$). The frequency in \cref{eq:plateshape} then depends on the indices only through $m^2+n^2$, which is unchanged when $m$ and $n$ are exchanged, so a mode and its transpose share a single natural frequency---for example $\omega_{(2,1)}=\omega_{(1,2)}$ because $2^2+1^2=1^2+2^2$. The two shapes remain physically distinct, one carrying its half-waves along $x$ and the other along $y$. Because they respond at the same frequency, however, any linear combination of them is itself a valid mode at that frequency: the two patterns are free to mix into an arbitrarily rotated pair. The modal decomposition is therefore non-unique within the degenerate subspace, and response data alone cannot single out the aligned pair. This difficulty has no analogue in the forward-PDE setting and is a stringent test for any learned modal method.

The boundary-value problems above yield the mode shapes and natural frequencies in closed form, but not the damping ratios. Damping does not follow from the undamped eigenproblem: $\zeta_i$ is set by the assumed dissipation model or identified from measured responses, as is standard in modal testing \citep{ewins2000modal,giurgiutiu2014shm}. Each mode is therefore characterized by the triple $(\phi_i,\omega_i,\zeta_i)$, of which the first two are known analytically for the four systems above and the third is prescribed when generating the response data (\cref{sec:setup}).

In practice the expansion \cref{eq:expansion} is truncated to the $r$ lowest modes, which carry the observable energy. Recovering the modal basis then means identifying $\{(\phi_i,\omega_i,\zeta_i)\}_{i=1}^{r}$ from a sampled response field $X\in\mathbb{R}^{T\times M}$ ($T$ time steps, $M$ spatial points), \emph{without} access to the analytical modes above. \Cref{fig:bc} summarizes the four configurations.

\subsection{ModalONet}
\label{sec:modal_lno}

\begin{figure*}[tbp]
  \vskip 0.1in
  \centering
    \resizebox{\textwidth}{!}{%
\begin{tikzpicture}[
  font=\sffamily,
  >={Stealth[length=2.3mm,width=1.9mm]},
  io/.style   ={circle,draw=edgecol,line width=.9pt,fill=ioblue,inner sep=1.2pt,minimum size=9mm,align=center},
  op/.style   ={circle,draw=edgecol,line width=.9pt,fill=oporange,inner sep=0pt,minimum size=7.5mm},
  tran/.style ={circle,draw=edgecol,line width=.9pt,fill=tranyellow,inner sep=0pt,minimum size=7mm},
  summ/.style ={circle,draw=edgecol,line width=.9pt,fill=summred,text=white,inner sep=0pt,minimum size=8.5mm},
  layer/.style={rounded corners=3pt,draw=edgecol,line width=1pt,fill=cream,align=center,inner sep=3pt},
  arr/.style  ={->,draw=edgecol,line width=.9pt},
  sub/.style  ={gray!55!black,font=\sffamily\scriptsize},
]
\node[inkcol,font=\sffamily\small\bfseries] at (-0.7,5.95) {(a)};
\node[io] (s) at (0,5.1) {$\mathbf{s}$};
\node[io] (t) at (0,3.3) {$t$};
\node[sub] at (0,4.55) {coordinate};
\node[sub] at (0,2.75) {time};
\node[op] (Pt) at (1.4,5.1) {$P$};
\node[op] (Pb) at (1.4,3.3) {$P$};
\node[layer,minimum width=28mm,minimum height=11mm] (trunk) at (4.1,5.1)
      {\textbf{Trunk net}\\[2pt]{\footnotesize separable $\psi_i(x)\chi_i(y)$}};
\node[layer,minimum width=28mm,minimum height=11mm] (lap) at (4.1,3.3)
      {\textbf{Laplace layer}\\[2pt]{\footnotesize (pole--residue)}};
\node[io] (phi) at (6.95,5.1) {$\phi(\mathbf{s})$};
\node[io] (q)   at (6.95,3.3) {$q(t)$};
\node[sub] at (6.95,5.72) {mode shapes};
\node[sub] at (6.95,2.75) {modal coordinates};
\node[summ] (mrg) at (8.5,4.2) {$\langle\cdot,\cdot\rangle$};
\node[io]   (xh)  at (9.9,4.2) {$\hat{x}$};
\node[sub]  at (9.15,3.52) {$\hat{x}(t,\mathbf{s})=\sum_i q_i\phi_i$};
\draw[arr] (s) -- (Pt);   \draw[arr] (t) -- (Pb);
\draw[arr] (Pt) -- (trunk); \draw[arr] (Pb) -- (lap);
\draw[arr] (trunk) -- (phi); \draw[arr] (lap) -- (q);
\draw[arr] (phi) to[out=-40,in=128] (mrg);
\draw[arr] (q)   to[out= 40,in=232] (mrg);
\draw[arr] (mrg) -- (xh);
\node[rounded corners=4pt,draw=inkcol,line width=1pt,fill=gray!4,
      minimum width=52mm,minimum height=30mm] (panel) at (13.1,4.2) {};
\node[inkcol,font=\sffamily\small\bfseries] at (13.1,5.30) {Modal basis \ (output)};
\draw[arr] (xh) -- (panel.west);
\foreach \cx/\fx/\lab/\fd in {11.70/{sin(deg(pi*x))*sin(deg(pi*y))}/{$\phi_1$}/{$\omega_1,\zeta_1$},
                              13.10/{sin(deg(2*pi*x))*sin(deg(pi*y))}/{$\phi_2$}/{$\omega_2,\zeta_2$},
                              14.50/{sin(deg(pi*x))*sin(deg(2*pi*y))}/{$\phi_3$}/{$\omega_3,\zeta_3$}}{
  \begin{axis}[modeaxis,anchor=center,at={(\cx cm,4.45cm)}]
    \addplot3[surf,shader=interp,draw=none,samples=52,domain=0:1,y domain=0:1] {\fx};
  \end{axis}
  \draw[edgecol,line width=.6pt] (\cx-0.60,3.85) rectangle (\cx+0.60,5.05);
  \node[inkcol,font=\sffamily\footnotesize] at (\cx,3.62) {\lab};
  \node[poleX,font=\sffamily\scriptsize]    at (\cx,3.26) {\fd};
}
\node[sub] at (13.1,2.84) {shapes $\phi_i$ + frequency $\omega_i$ + damping $\zeta_i$};
\node[inkcol,font=\sffamily\small\bfseries] at (-0.7,1.05) {(b)};
\def\bcx{7.45}
\node[rounded corners=4pt,draw=gray!55,line width=.9pt,fill=gray!3,
      minimum width=124mm,minimum height=27mm] (bframe) at (\bcx,-0.55) {};
\node[inkcol,font=\sffamily\footnotesize\itshape,anchor=north east] at ($(bframe.north east)+(-0.15,-0.1)$) {Laplace layer};
\draw[gray!55,dashed,line width=.8pt] (2.9,2.78) -- (bframe.north west);
\draw[gray!55,dashed,line width=.8pt] (5.3,2.78) -- (bframe.north east);
\node[io,minimum size=8mm] (vt) at (\bcx-4.45,-0.55) {$v(t)$};
\node[tran] (Lf) at (\bcx-2.85,-0.55) {$\mathcal{L}$};
\node[op,minimum size=7.5mm] (Vs) at (\bcx-1.35,0.05)  {$V(s)$};
\node[op,minimum size=7.5mm] (Ks) at (\bcx-1.35,-1.15) {$K_\phi(s)$};
\node[rounded corners=3pt,draw=polenode,line width=1pt,fill=blue!6,
      minimum width=26mm,minimum height=23mm] (pr) at (\bcx+0.85,-0.55) {};
\node[inkcol,font=\sffamily\footnotesize\bfseries] at (\bcx+0.85,0.40) {Pole--residue};
\begin{scope}[shift={(\bcx+0.85,-0.42)}]   
  \fill[blue!7] (-0.58,-0.40) rectangle (0,0.40);            
  \draw[->,gray!75,line width=.5pt] (-0.62,0)--(0.50,0) node[inkcol,font=\tiny,below=-1.8pt]{$\sigma$};
  \draw[->,gray!75,line width=.5pt] (0,-0.44)--(0,0.46) node[inkcol,font=\tiny,above right=-3pt]{$j\omega$};
  \foreach \px/\py in {-0.32/0.26,-0.32/-0.26,-0.19/0.13,-0.19/-0.13}
     \node[poleX,font=\footnotesize] at (\px,\py) {$\times$};
\end{scope}
\node[inkcol,font=\sffamily\scriptsize] at (\bcx+0.85,-1.20) {$\displaystyle\sum_i\frac{\beta_i}{s-\mu_i}$};
\node[tran] (Li) at (\bcx+3.15,-0.55) {$\mathcal{L}^{-1}$};
\node[io,minimum size=8mm] (qb) at (\bcx+4.55,-0.55) {$q(t)$};
\draw[arr] (vt) -- (Lf);
\draw[arr] (Lf) to[out=38,in=180]  (Vs);
\draw[arr] (Lf) to[out=-38,in=180] (Ks);
\draw[arr] (Vs) to[out=0,in=125] (pr.north west);
\draw[arr] (Ks) to[out=0,in=235] (pr.south west);
\draw[arr] (pr) -- (Li);
\draw[arr] (Li) -- (qb);

\end{tikzpicture}
}
    \caption{ModalONet architecture: (a)~The trunk maps spatial coordinates to continuous mode shapes $\phi_i(\bs)$, and the LNO branch maps time to modal coordinates $q_i(t)$ in pole--residue form, exposing the poles $\mu_i=-\sigma_i+j\omega_{d,i}$; their inner product reconstructs the field, from which the modal basis (mode shapes, frequencies, damping ratios) is read directly; (b)~The general LNO Laplace layer of \citet{cao2024laplace}, which maps an input $v(t)$ through a pole--residue transfer function $K_\phi(s)$; our branch specializes this layer to free decay, retaining the learnable poles and residues without a forcing input.\\}
    \label{fig:arch}
  \vskip -0.15in
\end{figure*}
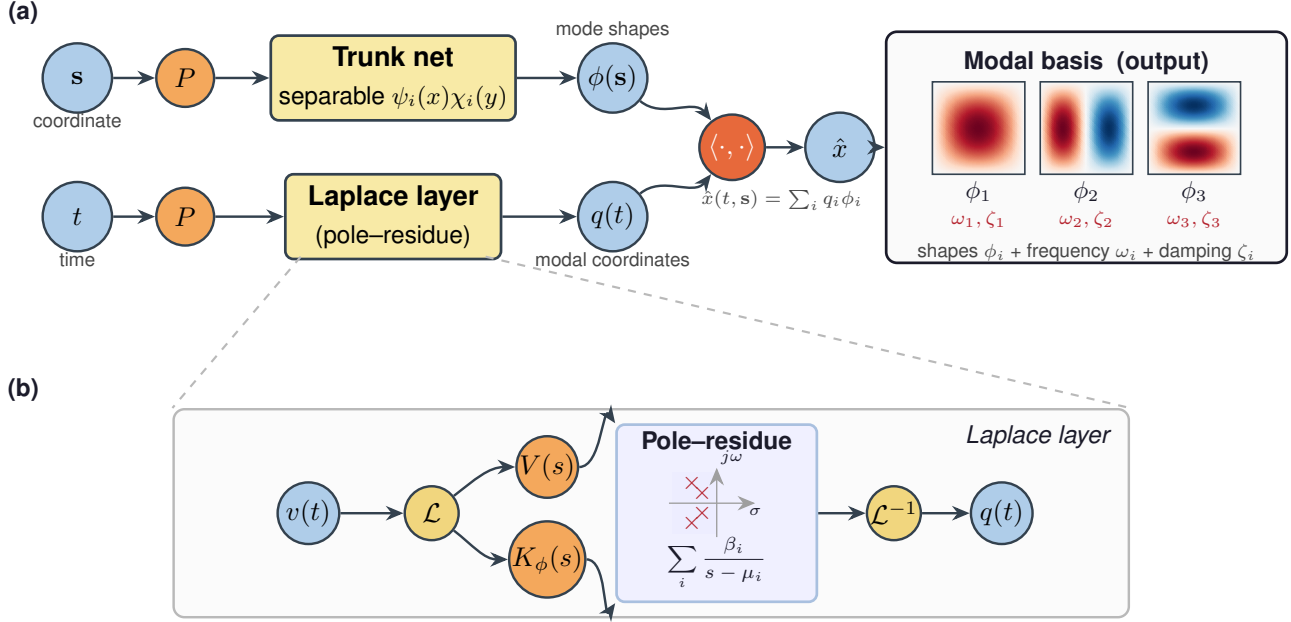

\textbf{From modal expansion to architecture.}
A DeepONet \citep{lu2021learning} represents an operator through a branch network $b$ and a trunk network $\tau$ whose outputs combine as a sum of products,
\begin{equation}
  \mathcal{G}(t,\bs)=\sum_{i=1}^{r} b_i(t)\,\tau_i(\bs),
  \label{eq:deeponet}
\end{equation}
a form that approximates continuous operators \citep{chen1995universal}. In this work we assign the trunk to the spatial coordinate and the branch to time. Comparing \cref{eq:deeponet} with the modal expansion \cref{eq:expansion} then reveals an exact correspondence: the trunk outputs $\tau_i(\bs)$ play the role of mode shapes, and the branch outputs $b_i(t)$ the role of modal coordinates. A DeepONet trained to reconstruct a response field is therefore already performing a decomposition of the form \cref{eq:expansion}; what it lacks is any reason for its factors to be the \emph{physical} modes rather than one of the infinitely many alternative factorizations of the same field. We therefore adopt the branch--trunk product as a structured representation of the response and constrain each factor to its physical counterpart. The trunk then recovers the mode shapes, and the branch the modal coordinates---and, through the poles that generate them, the natural frequencies and damping ratios. The reconstructed field is
\begin{equation}
  \xhat(t,\bs)=\sum_{i=1}^{r} q_i(t)\,\phi_i(\bs)
  = \big\langle\, q(t),\,\phi(\bs)\,\big\rangle,
  \label{eq:recon}
\end{equation}
where the mode shapes $\phi(\bs)$ come from the modal trunk and the modal coordinates $q(t)$ from the LNO branch, both described next and illustrated in \cref{fig:arch}.

\textbf{Modal trunk network.}
The trunk maps a spatial coordinate to the $r$ mode-shape values at that point, $\bs\mapsto\left(\phi_1(\bs),\dots,\phi_r(\bs)\right)$. Because it is a function of the coordinate rather than a vector tied to  sensor locations, the recovered shapes are continuous and mesh-free, and can be evaluated at any resolution. For 1-D domains the trunk is a single MLP. For 2-D domains we use a separable trunk, following Mandl et al., who factorize the trunk into per-coordinate sub-networks to break the curse of dimensionality in physics-informed operator learning
\citep{mandl2024separable}:
\begin{equation}
  \phi_i(x,y)=\psi_i(x)\,\chi_i(y),
  \label{eq:separable}
\end{equation}
where $\psi_i$ and $\chi_i$ are small MLPs acting on $x$ and $y$ separately. Separability mirrors the product structure of the analytical plate modes \cref{eq:plateshape}, and it also resolves the degeneracy identified in \cref{sec:modal}. A rotated mixture of two degenerate modes is in general not separable, so restricting the trunk to product form excludes such mixtures from its hypothesis space and steers the network toward the aligned pair.

\textbf{Laplace neural operator branch.}
The branch supplies the temporal factors. We follow the LNO \citep{cao2024laplace}, specialized here to free vibration. Rather than transforming a forcing input, the branch carries $r$ learnable poles $\mu_i$ and residues $\beta_i$, and evaluates
\begin{equation}
  q_i(t)=\operatorname{Re}\!\left\{\beta_i\,e^{\mu_i t}\right\},
  \qquad \mu_i=-\sigma_i + j\,\omega_{d,i},
  \label{eq:branch}
\end{equation}
which is the complex-exponential form of the classical free response \cref{eq:modalcoord}: the pole carries the decay rate $\sigma_i=\zeta_i\omega_i$ in its real part and the damped frequency $\omega_{d,i}$ in its imaginary part, while the residue $\beta_i$ encodes the amplitude and phase. This parameterization is what makes the modal quantities readable. A generic MLP branch can only approximate a decaying oscillation, and a Fourier branch presumes a periodicity that a decaying signal does not possess; the pole--residue form represents such a signal exactly and exposes the poles as native, trainable parameters \citep{cao2024laplace}. Since each pole governs the dynamics of one mode, the natural frequency and damping ratio follow from it in closed form, with no separate identification stage:
\begin{equation}
  \omega_i=\lvert\mu_i\rvert,
  \qquad
  \zeta_i=-\frac{\operatorname{Re}\mu_i}{\lvert\mu_i\rvert},
  \label{eq:extract}
\end{equation}
the natural frequency is the pole magnitude, the damping ratio is the cosine of the pole's angle to the negative real axis, and $\lvert\operatorname{Im}\mu_i\rvert$ recovers the damped frequency $\omega_{d,i}$. Two choices complete the branch. We parameterize $\sigma_i=e^{\eta_i}>0$, which pins every pole in the open left half-plane and so guarantees a decaying, stable response, and we initialize $\{\omega_i\}$ from an energy-ranked spectral decomposition of $X$ that remains robust for closely spaced or weakly excited modes (\cref{sec:impl}).

\subsection{Training objective}
\label{sec:loss}

The operator is trained on the response field alone; the analytical modes are never provided. The field is also consumed in raw form: $X$ enters the objective exactly as sampled, and no spectral or time--frequency transform is applied to it at any stage of training. This departs from prevailing identification practice, in which the network or estimator ingests a derived representation---a short-time Fourier or wavelet transform, a power spectral density, or a covariance-driven state-space realization---rather than the measured record. Such transforms impose their own resolution trade-offs and stationarity assumptions, and they attenuate the decay information on which damping estimation depends. The one spectral computation in the method, the singular value decomposition used to initialize the pole frequencies (\cref{sec:modal_lno}), runs once before training and never enters the loss, which is evaluated entirely on time-domain samples.

The objective combines reconstruction with constraints that make the latent factors a modal basis,
\begin{equation}
  \mathcal{L}=\mathcal{L}_{\mathrm{rec}}
  +\lambda_o\mathcal{L}_{\mathrm{orth}}
  +\lambda_n\mathcal{L}_{\mathrm{norm}}
  +\lambda_t\mathcal{L}_{\mathrm{proj}}.
  \label{eq:loss}
\end{equation}
The reconstruction term is the field misfit $\mathcal{L}_{\mathrm{rec}}=\lVert\hat{X}-X\rVert_F^2$, which alone would be satisfied by any factorization of the response and therefore cannot single out the physical modes; the remaining three terms supply that selection.

Let $\Phi=[\phi_1\ \cdots\ \phi_r]\in\mathbb{R}^{M\times r}$ denote the trunk-output matrix and $G=\Phi^\top\mathrm{diag}(\mathrm{d}\bs)\,\Phi$ its discrete Gram matrix under the inner product \cref{eq:ortho}. Orthogonality and normalization penalties then enforce an orthonormal spatial basis,
\begin{equation}
  \mathcal{L}_{\mathrm{orth}}=\sum_{i\neq j}G_{ij}^2,
  \qquad
  \mathcal{L}_{\mathrm{norm}}=\sum_i (G_{ii}-1)^2.
  \label{eq:gramloss}
\end{equation}
Together these drive $G\rightarrow I$, realizing \cref{eq:ortho} discretely: the off-diagonal term keeps distinct mode shapes from collapsing onto one another, while the diagonal term fixes the scale that the eigenproblem itself leaves arbitrary (\cref{sec:modal}).

The temporal projection consistency term ties each mode shape to the data through its own time signature, energy-normalized,
\begin{equation}
  \mathcal{L}_{\mathrm{proj}}=\sum_i
  \Big\lVert \phi_i - X^\top\!\big(q_i/\lVert q_i\rVert_2^2\big)\Big\rVert_2^2.
  \label{eq:proj}
\end{equation}
Dividing by $\lVert q_i\rVert_2^2$ rescales each mode's projection by its own temporal energy. Without it, a briefly excited, fast-decaying mode contributes little to the time-averaged reconstruction and drifts toward the dominant modes; with it, every mode receives a well-conditioned spatial gradient regardless of how strongly it was excited. The weights $\lambda_o,\lambda_n,\lambda_t$ are fixed across all experiments; architectures and optimization settings are collected in \cref{sec:impl}.

\begin{figure*}[tb]
  \vskip 0.1in
  \centering
    \resizebox{\textwidth}{!}{%
\begin{tikzpicture}[
  font=\sffamily,
  >={Stealth[length=2.5mm,width=2mm]},
  box/.style={rounded corners=3pt,draw=nodeln,line width=.9pt,fill=white,align=center,
              minimum height=12mm,font=\sffamily\small,text=ink,inner sep=3pt},
  data/.style={box,fill=datafill,draw=databd},
  model/.style={box,minimum width=34mm},
  loss/.style={box,draw=lossbd,fill=lossbg,minimum width=24mm},
  acc/.style={box,fill=accfill,draw=accfill,text=white,line width=1pt,minimum width=30mm},
  fl/.style={->,draw=edgec,line width=1pt},
  el/.style={font=\sffamily\scriptsize,text=edgec,inner sep=1.5pt,fill=white,align=center},
  gl/.style={font=\sffamily\small\bfseries,text=ink},
]
\node[data]  (X)     at (0,0)    {$X(t,\mathbf{s})$};
\node[model] (M)     at (3.1,0)  {\textbf{ModalONet}\\[1pt]{\footnotesize trunk $\phi_i(\mathbf{s})$\ $\cdot$\ branch $q_i(t)$}};
\node[loss]  (L)     at (6.6,0)  {Composite\\loss $\mathcal{L}$};
\node[box]   (P)     at (10.8,0) {Read poles\\[1pt]{\footnotesize$\mu_i\!\to\!(\omega_i,\zeta_i)$}};
\node[box]   (Mt)    at (13.7,0) {Match, align,\\[1pt]{\footnotesize refine $\zeta_i$}};
\node[acc]   (B)     at (16.7,0) {\textbf{Modal basis}\\[1pt]{\footnotesize$\{\phi_i,\omega_i,\zeta_i\}$}};

\begin{scope}[on background layer]
  \node[rounded corners=5pt,fill=trnbg,draw=trnbd,line width=.8pt,
        fit=(X)(M)(L),inner xsep=5mm,inner ysep=10mm](gT){};
  \node[rounded corners=5pt,fill=extbg,draw=extbd,line width=.8pt,
        fit=(P)(Mt)(B),inner xsep=5mm,inner ysep=10mm](gE){};
\end{scope}
\node[gl,anchor=north west] at ([shift={(2mm,-1.6mm)}]gT.north west){Training loop\ \small(unsupervised)};
\node[gl,anchor=north west] at ([shift={(2mm,-1.6mm)}]gE.north west){Modal extraction};

\draw[fl] (X) -- (M);
\draw[fl] (M) -- (L) node[el,midway,yshift=0.5pt]{$\hat{X}$};
\draw[fl,dashed] (L.south) |- ([yshift=-7mm]$(M.south)!0.5!(L.south)$) -| (M.south)
       node[el,pos=0.5]{Adam update};
\draw[fl,line width=1.1pt] (L) -- (P) node[el,above,midway,yshift=1pt]{after\\convergence};
\draw[fl] (P) -- (Mt) node[el,midway]{MAC};
\draw[fl] (Mt) -- (B);
\end{tikzpicture}
}
    \caption{Method overview. The response field drives the trunk and LNO branch; their reconstruction is scored by the composite loss function in \cref{eq:loss} and optimized over the training loop. After convergence, poles are read out, MAC-based Hungarian assignment and sign alignment order the modes, and a log-envelope regression refines damping ratios, yielding the modal basis $\{\phi_i,\omega_i,\zeta_i\}$.\\}
    \label{fig:flow}
  \vskip -0.2in
\end{figure*}
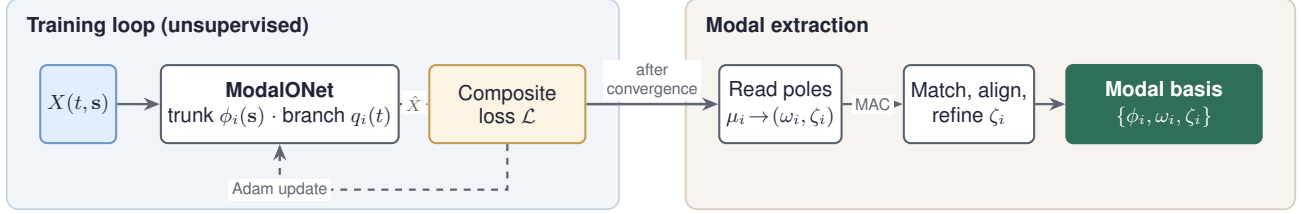

\textbf{Degenerate-mode resolution.}

As described in \cref{sec:modal}, a square plate ($a=b$) exhibits degenerate modes: the pair $(2,1)/(1,2)$ shares a single natural frequency, and any rotation of the two is an equally valid pair at that frequency. Degeneracy therefore poses two distinct problems, one temporal and one spatial. The temporal problem is a genuine identifiability limit: two damped sinusoids at a common frequency sum to a signal containing a \emph{single} frequency, so no amount of response data can apportion a frequency split within the pair. Left free, the two poles resolve this indeterminacy arbitrarily and drift to two different, incorrect values. We therefore \emph{tie} degenerate poles: modes whose SVD-initialized frequencies coincide within $2\%$ share one learnable frequency, while damping remains per-mode, since degenerate modes may dissipate differently. The grouping is detected from the data and injects only the symmetry fact that the pair shares a frequency, never that frequency's value; it thus removes a degree of freedom the data cannot constrain rather than adding supervision. Systems without degeneracy form singleton groups and are unaffected.

The spatial problem---the freedom to return a rotated mixture---is already prevented by the separable trunk \cref{eq:separable}. What remains after training is only an arbitrary ordering and sign of the recovered modes, which we remove by pairing learned to analytical modes through the Hungarian assignment maximizing
$\sum_i \mathrm{MAC}_{\pi(i),i}$, where
\begin{equation}
  \mathrm{MAC}_{ij}
  =\frac{\big\langle \phi_i,\phi_j^{\mathrm{ref}}\big\rangle^{2}}
        {\lVert\phi_i\rVert^{2}\,\lVert\phi_j^{\mathrm{ref}}\rVert^{2}}
  \label{eq:mac}
\end{equation}
is the modal assurance criterion under the inner product \cref{eq:ortho}, and by aligning signs with the corresponding $L^2$ cosine. MAC is the standard mode-shape agreement metric of experimental modal analysis: it is bounded in $[0,1]$ and sign-invariant, and, unlike Pearson correlation---whose mean removal manufactures spurious cross-correlation between $L^2$-orthogonal modes with nonzero domain means---it is mean-free, so it respects exactly the orthogonality enforced by \cref{eq:gramloss}. This matching step affects only labelling and orientation; it does not alter the recovered shapes, frequencies, or damping.

\begin{algorithm}[b]
  \caption{ModalONet: training and extraction}
  \label{alg:train}
  \begin{algorithmic}
    \STATE {\bfseries Input:} response field $X\in\mathbb{R}^{T\times M}$; rank $r$;
           weights $\lambda_o,\lambda_n,\lambda_t$
    \STATE Initialize $\{\omega_i\}$ from the SVD of $X$; initialize trunk and branch parameters
    \STATE tie near-coincident initial frequencies (within $2\%$) to one shared $\omega$ per degenerate group
    \FOR{$k=1$ {\bfseries to} $N_{\mathrm{epochs}}$}
      \STATE $\phi\leftarrow$ trunk$(\bs)$;\quad $q\leftarrow$ LNO branch$(t)$
      \STATE $\hat{X}\leftarrow q\,\phi^\top$
      \STATE form $G$; evaluate $\mathcal{L}$ via
             \cref{eq:loss,eq:gramloss,eq:proj}
      \STATE Adam step on trunk and branch parameters
    \ENDFOR
    \STATE read poles $\mu_i$: $\omega_i\!=\!|\mu_i|$,
           $\zeta_i\!=\!-\mathrm{Re}\,\mu_i/|\mu_i|$
    \STATE pair to references by Hungarian assignment maximizing
           $\sum_i \mathrm{MAC}_{\pi(i),i}$ \cref{eq:mac}; align signs via the $L^2$ cosine
    \STATE refine $\zeta_i$ by SNR-weighted log-envelope regression on the
           modal projections \cref{eq:modalproj}
    \STATE {\bfseries Output:} modal basis $\{(\phi_i,\omega_i,\zeta_i)\}_{i=1}^r$
  \end{algorithmic}
\end{algorithm}

\textbf{Damping refinement.}
The reconstruction loss constrains each decay rate $\sigma_i$ only weakly, because the residue $\beta_i$ can absorb amplitude changes over the fitted window; pole damping is therefore biased low, most severely for the degenerate pair. We refine damping post hoc from quantities the model does pin down. Since the learned shapes are near-orthonormal, each modal coordinate is directly observable by spatial projection,
\begin{equation}
  y_i(t)=\frac{\big\langle X(t,\cdot),\,\phi_i\big\rangle}{\lVert\phi_i\rVert^{2}},
  \label{eq:modalproj}
\end{equation}
whose analytic (Hilbert) envelope decays as $e^{-\zeta_i\omega_i t}$. A weighted linear regression of the log-envelope against time over the above-noise-floor window---with weights proportional to the squared envelope, since the log of a noisy envelope is heteroscedastic---gives $\zeta_i=-\mathrm{slope}/\omega_i$, a classical logarithmic-decrement estimate. Damping of degenerate modes remains identifiable in this way even though their frequency split is not: the pair shares a frequency, but their shapes are orthogonal, so the projection \cref{eq:modalproj} separates their envelopes. Like the rest of the pipeline, the refinement consumes only the raw field and quantities already learned. \Cref{alg:train} summarizes training and extraction, and \cref{fig:flow} depicts the pipeline.


\section{Numerical Experiments}
\label{sec:experiments}

\subsection{Setup}
\label{sec:setup}

We evaluate four systems, each with $r=3$ target modes: simply supported and cantilever Euler--Bernoulli beams, and SSSS Kirchhoff plates that are rectangular ($a{=}1.5$, $b{=}1.0$) or square ($a{=}b{=}1.0$). For the plates the non-dimensional stiffness is set to $\sqrt{D/\rho h}=0.2$; this fixes only the overall frequency scale, while the mode-frequency \emph{ratios}, the scale-invariant physical quantities of \cref{eq:plateshape}, are unaffected.

Training data are modal-superposition response fields built from the analytical bases of \cref{sec:modal}: free-decay responses from random modal initial conditions, with per-mode amplitudes drawn uniformly from $[0.7,1.3]$ and phases from $[0,2\pi)$, true damping ratios $\zeta=(0.020,0.030,0.025)$ for modes 1--3 of every system, and beam properties non-dimensionalized to $EI/m=1$. Fields are sampled at $T=2{,}000$ instants with step $\Delta t=0.005$ (a $10$\,s record) on spatial grids of $M=30$ (SS beam), $M=40$ (cantilever), and $15\times 15$ (rectangular and square plate) points. We corrupt the field with zero-mean i.i.d.\ Gaussian measurement noise of standard deviation $\sigma_\varepsilon=0.01$ in these non-dimensional units---about $1\%$ of the fundamental's peak modal amplitude, a signal-to-noise ratio of ${\approx}30$; the network observes only the noisy field $X$, never the analytical modes.

Every system is trained with the settings of \cref{sec:impl}. We report the modal assurance criterion \cref{eq:mac} between each learned and analytical mode shape, the relative errors of the recovered natural frequencies and refined damping ratios, and the Gram matrix of the learned shapes. Synthetic data are a deliberate choice: because the generating modal basis is known in closed form, the recovered basis can be verified against an exact ground truth, and as no prior neural operator outputs a modal basis, this analytical eigen solution is the natural reference.

\section{Results and Discussion}
\label{sec:results}

\subsection{Implementation}
\label{sec:impl}

The 1-D trunk is an MLP with three hidden layers of width $64$ and $\tanh$ activations. In the separable 2-D trunk \cref{eq:separable}, each per-coordinate sub-network $\psi_i,\chi_i$ has two hidden layers of width $64$. The branch holds $r$ learnable poles and complex residues, with the decay rates parameterized as $\sigma_i=e^{\eta_i}$ to keep every pole in the left half-plane.
Pole frequencies are initialized from a singular value decomposition of the response field: the first $r$ temporal singular vectors are extracted, and the dominant peak of each of their spectra provides one frequency estimate. Initial frequencies that coincide within $2\%$ are grouped and tied to a single shared parameter (\cref{sec:loss}). Trunk and branch are then trained jointly with Adam under a cosine-annealed learning rate (initial value $1.5\times10^{-3}$) for $15{,}000$ iterations, with loss weights $\lambda_o=20$, $\lambda_n=2$, and $\lambda_t=10$ held fixed across every system reported in this work.

The damping refinement of \cref{sec:loss} proceeds as follows. The analytic envelope of each modal projection \cref{eq:modalproj} is smoothed; samples below four times the projected noise floor---estimated from the reconstruction residual---are discarded, as are the first and last $3\%$ of the record, where the Hilbert transform is unreliable; and a weighted least-squares line with weights proportional to the squared envelope gives $\hat{\zeta}_i=-\mathrm{slope}/\omega_i$. All runs use fixed random seeds and complete in minutes on a single CPU. Code and configurations will be released.

\subsection{Mode-shape recovery}
\label{sec:shapes}

\begin{figure*}[tbp]
  \vskip 0.1in
  \centering
    \includegraphics[width=0.8\textwidth]{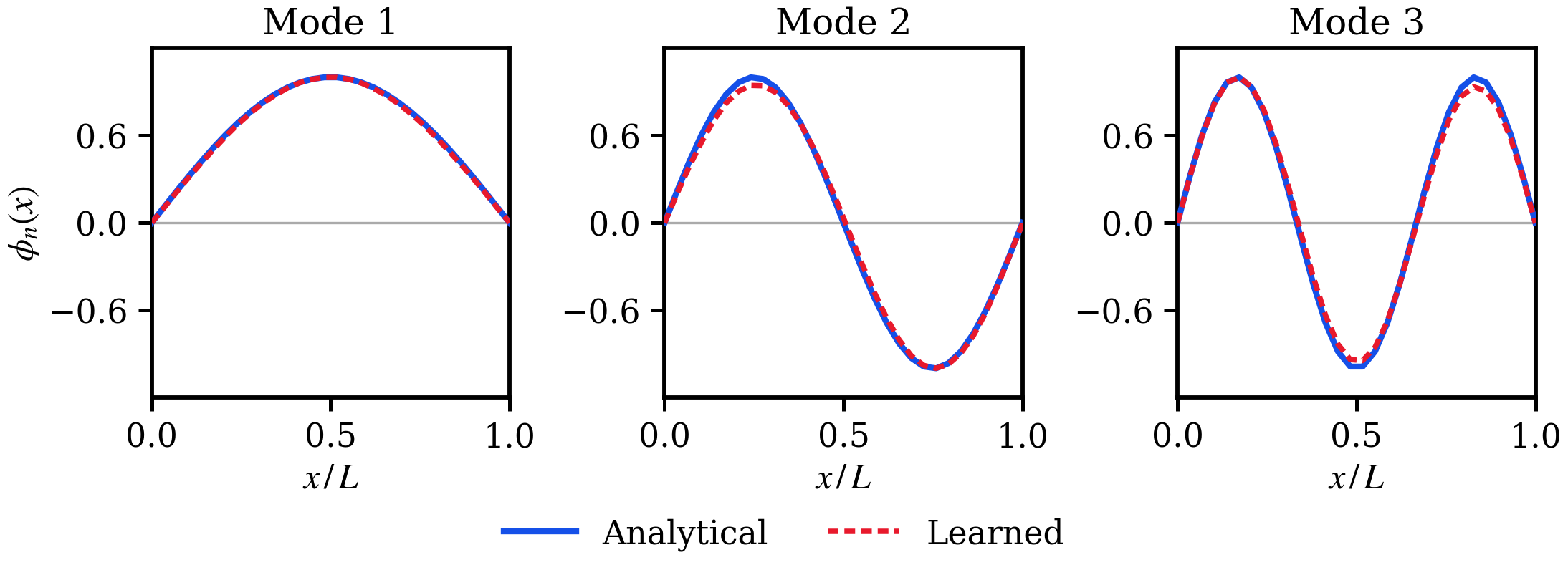}
    \caption{Simply supported beam: analytical (solid) versus learned (dashed) mode shapes for the first three modes.}
    \label{fig:beam}
  \vskip -0.15in
\end{figure*}

\begin{figure*}[tbp]
  \vskip 0.1in
  \centering
    \includegraphics[width=0.8\textwidth]{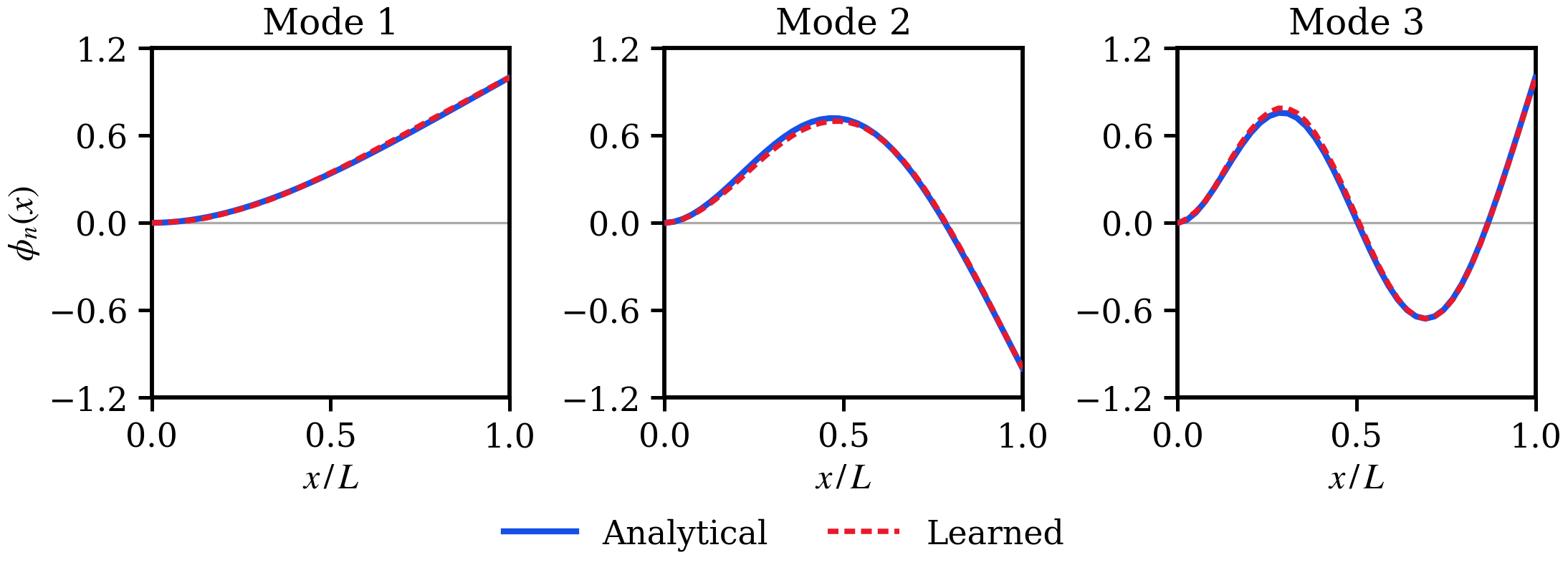}
    \caption{Cantilever beam: analytical versus learned mode shapes. The non-sinusoidal clamped--free shapes of \cref{eq:cantshape} are recovered at MAC $\geq 0.999$.\\}
    \label{fig:cant}
  \vskip -0.15in
\end{figure*}

Mode shapes are the most demanding of the three modal quantities, since each is a function over the domain rather than a scalar. \Cref{fig:beam} shows the simply supported beam, whose learned shapes match the analytical $\sin(n\pi x/l)$ basis of \cref{eq:ssbeam} at MAC $\geq 0.998$ for all three modes. The cantilever in \cref{fig:cant} is a stricter test, because its shapes are the non-sinusoidal $\cosh{-}\cos{-}\beta(\sinh{-}\sin)$ forms of \cref{eq:cantshape}; they are recovered at MAC $\geq 0.999$, indicating that the trunk is not biased toward sinusoids by its architecture.

\begin{figure}

\setlength{\unitlength}{1pt}%
\begin{picture}(234,130)
    \put(0,0){ \includegraphics[width=1\columnwidth]{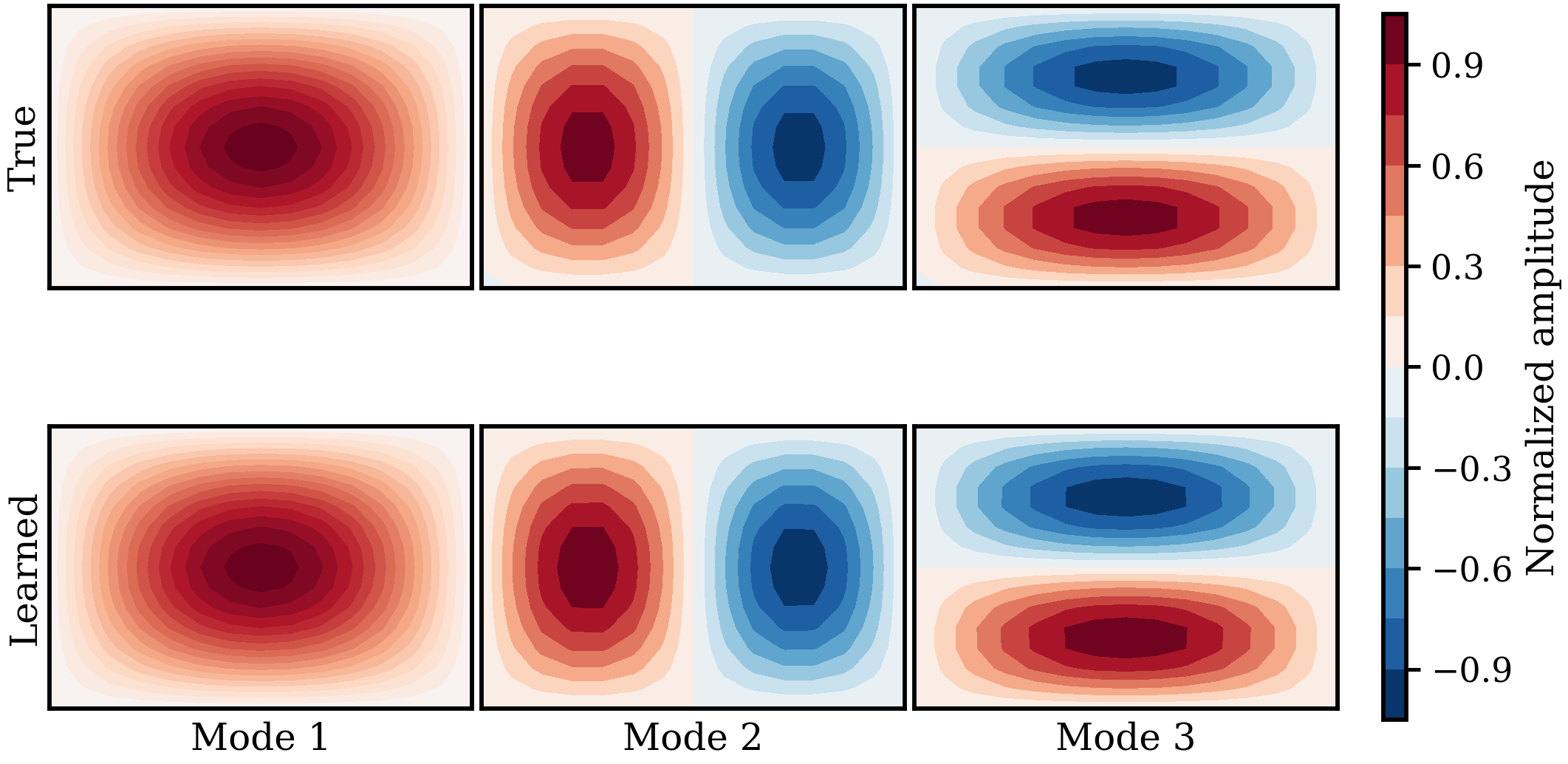}}
    \put(0,-170){\includegraphics[width=1\columnwidth]{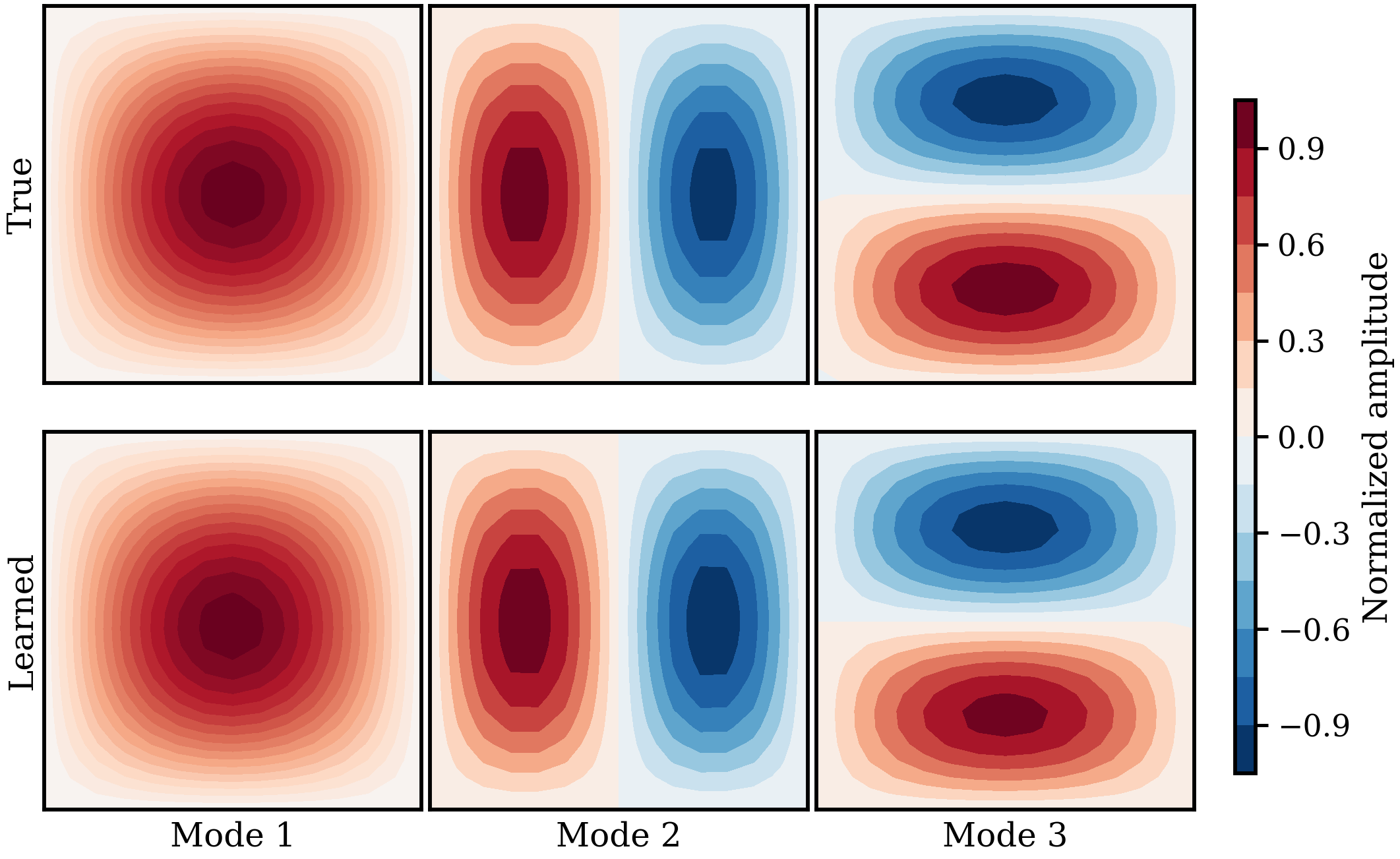}}
    \put(0,118){\large \textbf{(a): Rectangular plate ($a{=}1.5,b{=}1.0$)}}
    \put(0,-20){ \large \textbf{(b): Square plate ($a{=}b{=}1.0$)}}
    \end{picture}

   \vspace{6cm}\caption{Plate mode shapes: analytical (top row of each panel) versus learned (bottom row). The separable trunk separates the degenerate square-plate pair rather than returning a mixture; Hungarian assignment fixes its order; the pair $(2,1)/(1,2)$ is degenerate.}
\label{fig:plates}%
\end{figure}

\Cref{fig:plates} turns to the two-dimensional cases. On the rectangular plate (\cref{fig:plates}a), the three non-degenerate modes $(1,1)$, $(2,1)$, and $(1,2)$ are recovered at MAC $=1.000$ to four decimals. The square plate (\cref{fig:plates}b) is the degenerate case of \cref{sec:modal}: the separable trunk returns the two members of the pair $(2,1)/(1,2)$ individually rather than as a mixture, the assignment step supplies their order, and the MAC values are $0.999$, $1.000$, and $0.999$. \Cref{tab:corr} collects frequencies and damping ratios for all four systems.

\subsection{Orthonormality of the learned basis}
\label{sec:gram}

\begin{figure*}[tbp]
  \vskip 0.1in
  \centering
    \includegraphics[width=\textwidth]{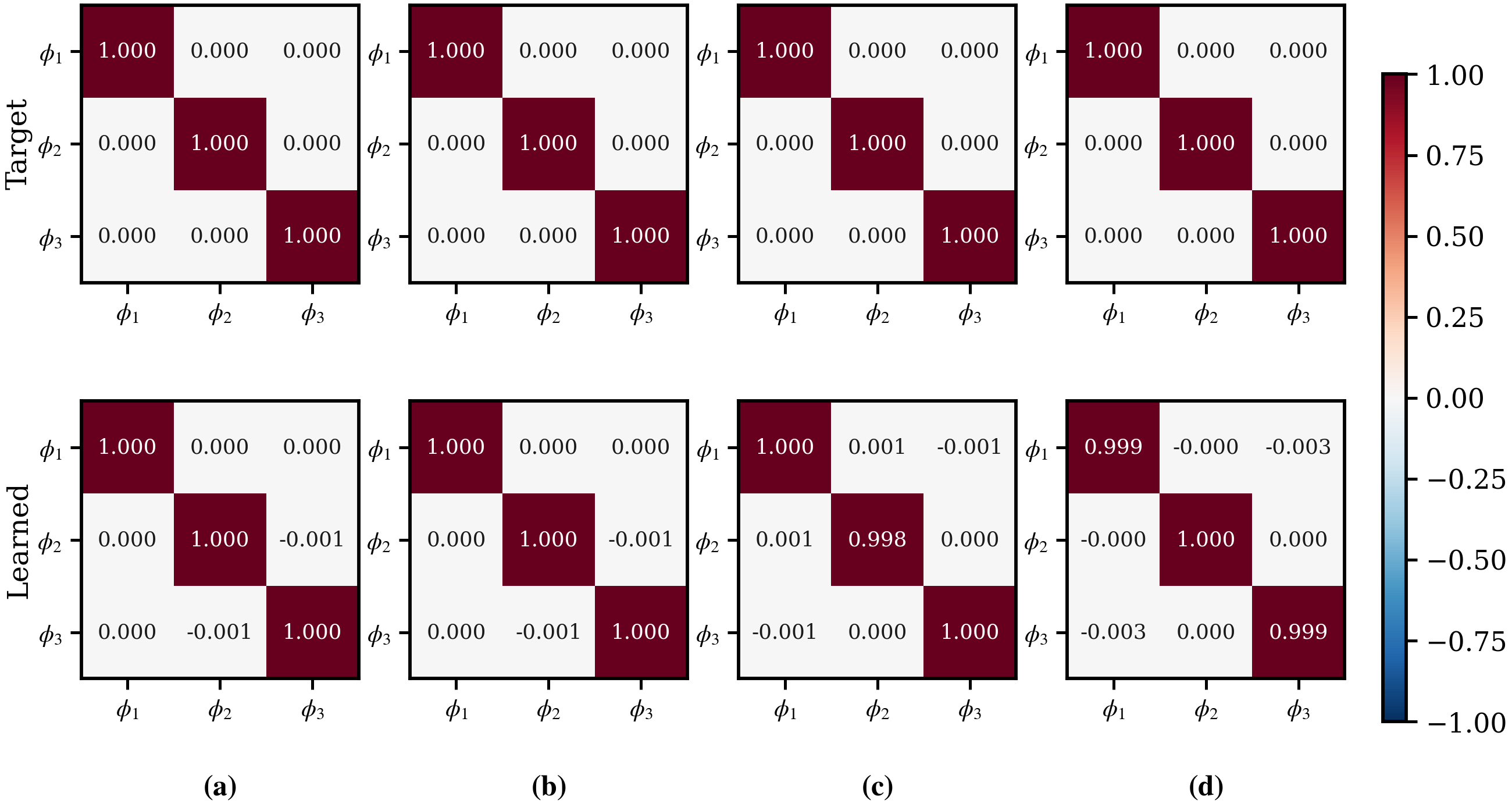}
    \caption{Gram matrices of the target (top row) and learned (bottom row) mode shapes for (a) a simply supported Euler--Bernoulli beam, (b) a cantilever beam, (c) a rectangular Kirchhoff plate, and (d) a square Kirchhoff plate. The identity Gram matrices of the target modes confirm orthonormality, while the learned Gram matrices demonstrate that ModalONet preserves this property with only negligible off-diagonal terms, even in the presence of degenerate modes.}
    \label{fig:gram}
  \vskip -0.15in
\end{figure*}

Accurate shapes do not by themselves guarantee a well-formed basis, and the Gram penalties \cref{eq:gramloss} act only on the trunk output, so whether they succeed is an empirical question. \Cref{fig:gram} compares, for every system, the target Gram matrix (the identity) with the learned one. The learned matrices are diagonal to within a few thousandths: the recovered shapes are individually normalized and mutually orthogonal, not merely a well-fitting spanning set.

\begin{figure*}[tbp]
  \vskip 0.1in
  \centering
    \includegraphics[width=\textwidth]{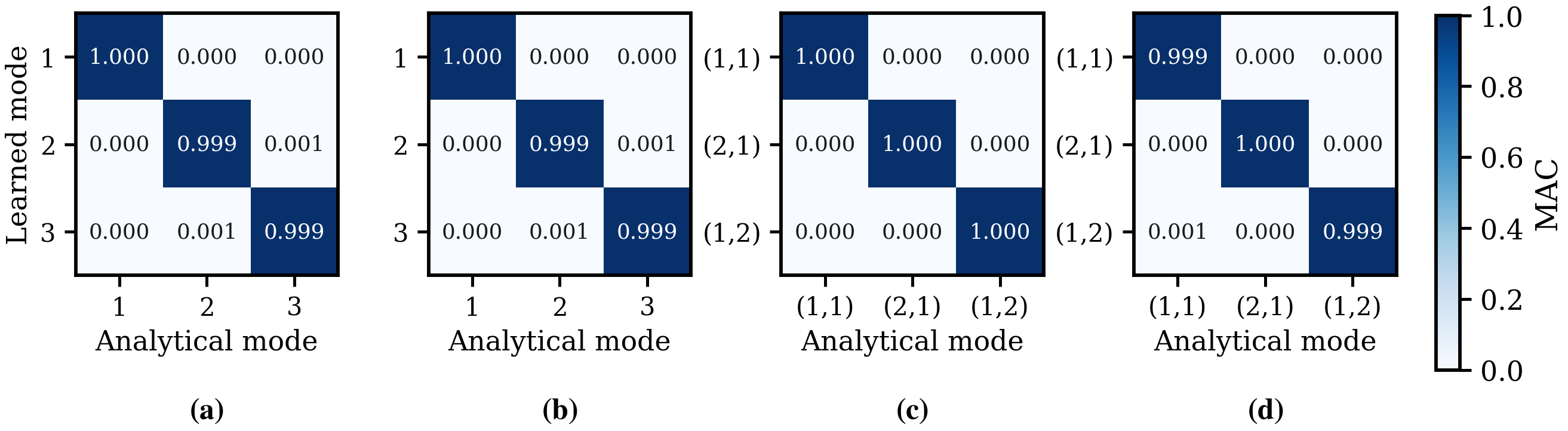}
    \caption{Modal assurance criterion (MAC) matrices between the learned and analytical mode shapes for (a) a simply supported beam, (b) a cantilever beam, (c) a rectangular Kirchhoff plate, and (d) a square Kirchhoff plate. Strong diagonal entries and negligible off-diagonal values indicate accurate one-to-one correspondence between the learned and analytical modes, including the degenerate square-plate case. In contrast to the Gram matrices in \cref{fig:gram}, which assess the internal orthonormality of the learned modal basis, the MAC matrices quantify agreement between the learned and analytical mode shapes.\\}
    \label{fig:corr}
  \vskip -0.15in
\end{figure*}

The Gram matrix is, however, an internal check---a rotated set of orthonormal modes would satisfy it equally well. \Cref{fig:corr} therefore reports the learned-versus-analytical MAC matrices \cref{eq:mac}, which measure agreement with ground truth: a strong diagonal with near-zero off-diagonal entries confirms one-to-one alignment and no mode mixing. The two matrices are complementary---one certifies that the basis is well formed, the other that it is the \emph{correct} basis---and \cref{sec:ablation} exhibits a case in which the first holds while the second fails.

\subsection{Frequency and damping ratio recovery}
\label{sec:freq}

\begin{figure*}[tbp]
  \vskip 0.1in
  \centering
    \includegraphics[width=0.85\textwidth]{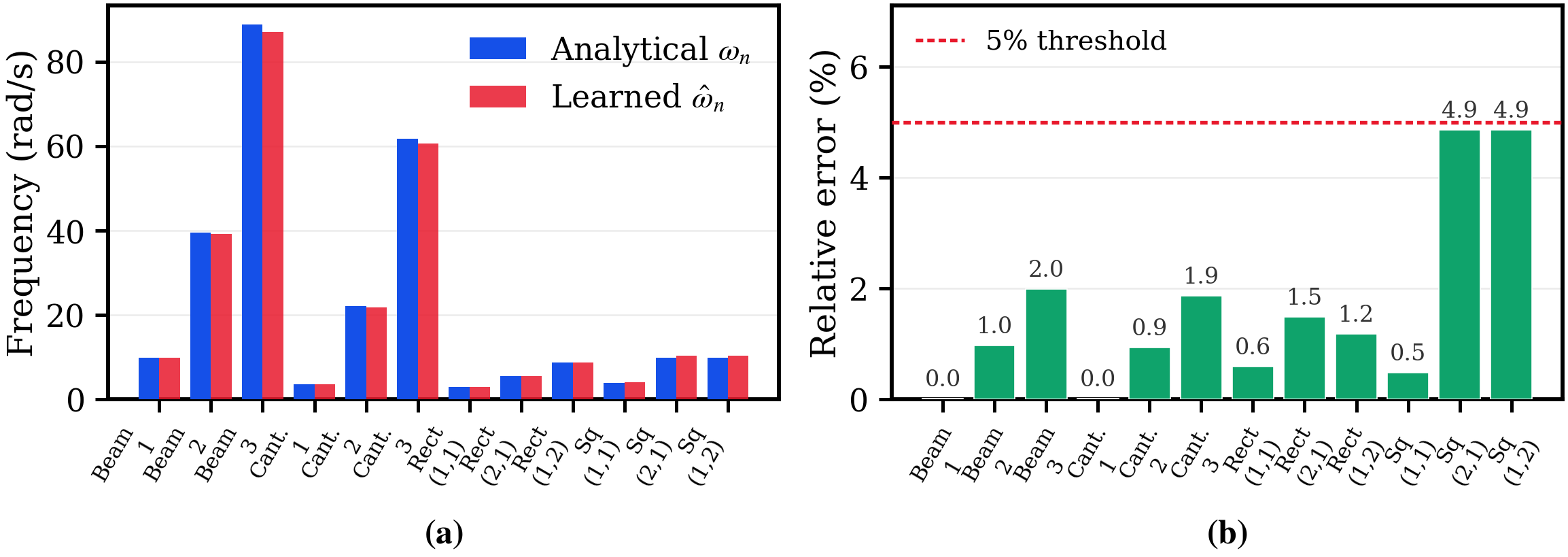}
    \caption{Natural-frequency recovery across the four structural systems: (a) analytical and ModalONet-predicted natural frequencies for all twelve modes; (b) corresponding per-mode relative errors; all predicted frequencies satisfy the 5\% error threshold (dashed line), with the largest error of 4.9\% occurring for the degenerate (1,2) and (2,1) modes of the square Kirchhoff plate.}
    \label{fig:freq}
  \vskip -0.15in
\end{figure*}

The natural frequencies are read from the learned poles via \cref{eq:extract}. \Cref{fig:freq} compares analytical and learned frequencies for all twelve modes and reports the per-mode relative error. Every mode is recovered within $5\%$, and the non-degenerate modes within $2\%$. The square-plate pair is represented by a single tied pole (\cref{sec:loss}) and recovers its shared frequency to $4.9\%$. With two free poles the branch resolves the unidentifiable split arbitrarily, reporting two different and incorrect values ($11.1\%$ and $4.9\%$ in an untied run), the expected signature of a degree of freedom the data cannot constrain. The residual error on the shared value reflects the limited observability of a quickly decaying, equal-frequency pair under fixed measurement noise, not a split.

The damping ratios read directly from the poles are markedly less reliable. The reconstruction loss is nearly insensitive to the decay rates $\sigma_i$, because the residues absorb amplitude over the fitted window, so pole damping is biased low---mildly for well-observed beam modes, and severely for the degenerate pair, whose worst raw error reaches $-97\%$.

The log-envelope refinement of \cref{sec:loss} corrects this from the data. The refined ratios in \cref{tab:corr} have a worst-case error of $-6.5\%$ (square plate, mode $(1,2)$), with nine of twelve modes within $2\%$ of the target, and the refinement recovers \emph{distinct} damping for the two degenerate modes---identifiable through their orthogonal shapes via \cref{eq:modalproj} even though their frequency split is not. For that pair, most of the residual is inherited rather than intrinsic to the envelope fit: the regression estimates the product $\zeta_i\omega_i$, so dividing by a shared frequency that sits $4.9\%$ high biases $\zeta_i$ low by $4.6\%$---almost exactly the error observed for $(2,1)$ ($-4.6\%$), and the bulk of that for $(1,2)$ ($-6.5\%$). The square fundamental is a separate case: its frequency is accurate to $0.5\%$, so this mechanism accounts for only $-0.5\%$ of its $-3.8\%$ deviation, which we attribute instead to the square plate's larger overall reconstruction residual, roughly an order of magnitude above the other systems.

\begin{figure*}[tbp]
  \vskip 0.1in
  \centering
    \includegraphics[width=0.9\textwidth]{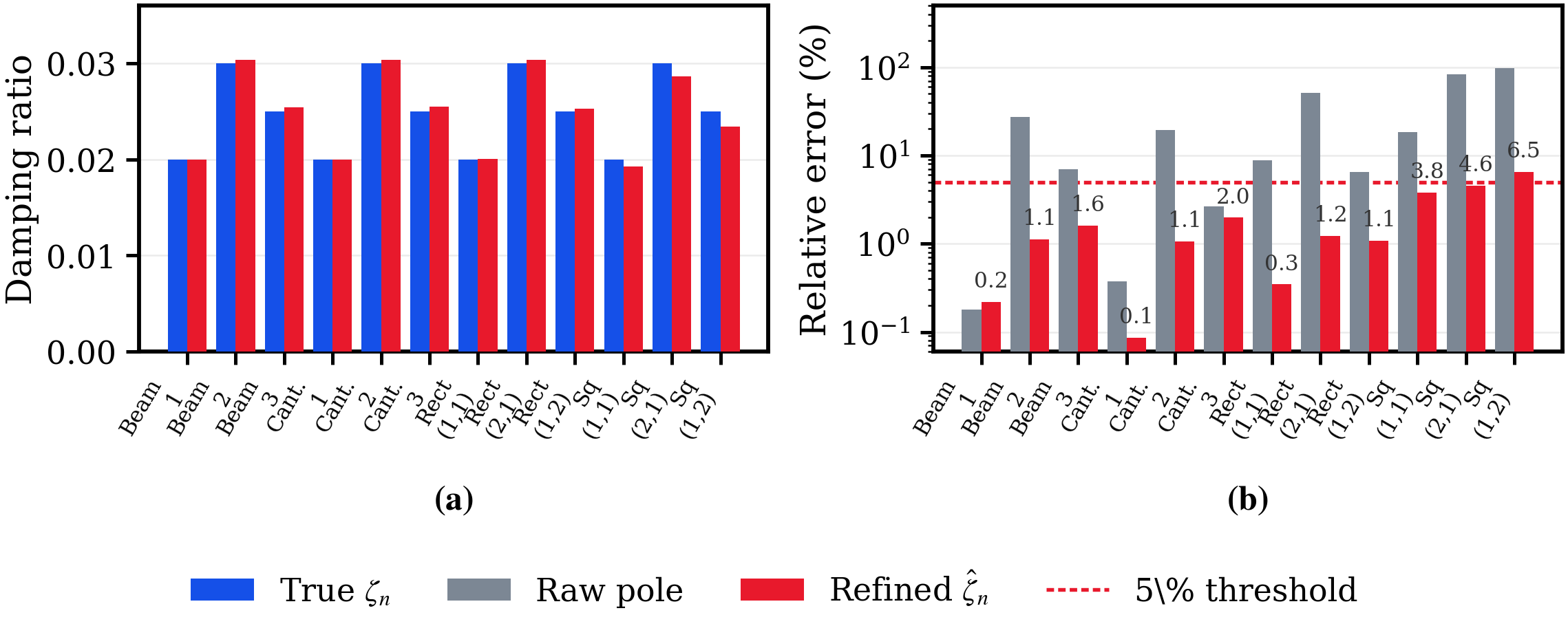}
    \caption{Damping-ratio recovery: (a) analytical (true) damping ratios and the log-envelope–refined ModalONet estimates for all twelve modes; (b) relative error (log scale) of the raw pole estimates (gray) and the refined estimates (red), with the dashed line indicating the 5\% error threshold; the log-envelope refinement reduces the worst-case error from 97\% to 6.5\%, with nine of the twelve modes achieving errors below 2\%.}
    \label{fig:damprec}
  \vskip -0.15in
\end{figure*}

\begin{figure*}[tbp]
  \vskip 0.1in
  \centering
    \includegraphics[width=0.9\textwidth]{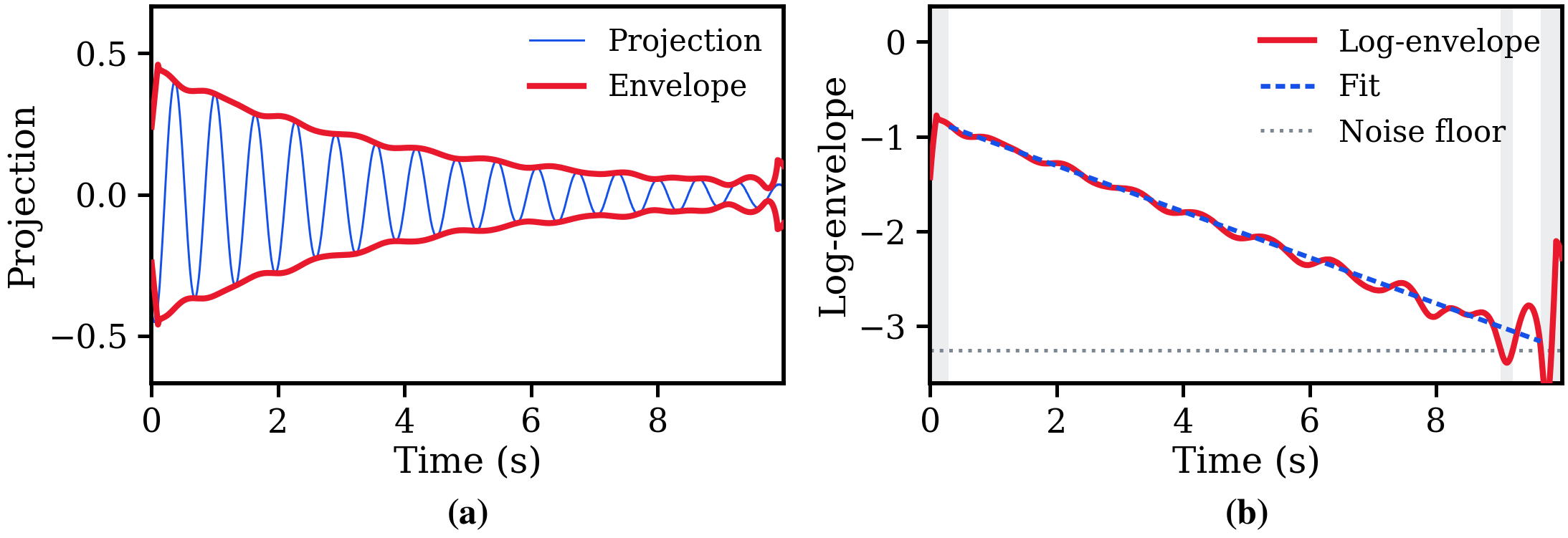}
    \caption{Damping-ratio refinement for the degenerate $(1,2)$ mode of the square Kirchhoff plate: (a) Modal projection $y_i(t)$ \cref{eq:modalproj} (blue) and its Hilbert envelope (red); (b) Logarithm of the envelope (red) together with the SNR-weighted least-squares linear fit (blue dashed); Shaded regions indicate samples excluded from the regression because of edge effects or amplitudes below the $4\sigma_\varepsilon$ projected noise floor (gray dotted line). The fitted slope yields the refined damping-ratio estimate $\hat\zeta=-\mathrm{slope}/\hat\omega=0.023$.}
    \label{fig:dampenv}
  \vskip -0.15in
\end{figure*}

\Cref{fig:damprec} summarizes damping recovery across all twelve modes: the refined ratios track the targets closely (panel~a), and the per-mode error panel (panel~b) contrasts the raw pole estimates---which range from near-exact on the well-observed fundamentals to $-97\%$ on the square-plate degenerate pair---with the refinement, which brings every mode to within $6.5\%$. \Cref{fig:dampenv} illustrates the mechanism on the hardest case (square plate, mode $(1,2)$): the modal projection \cref{eq:modalproj} and its Hilbert envelope (panel~a), and the log-envelope with the SNR-weighted fit over the retained high-signal window (panel~b), whose slope yields $\hat\zeta=0.023$ against a true $0.025$.

\subsection{Training behavior}
\label{sec:training}

\begin{figure*}[tbp]
  \vskip 0.1in
  \centering
    \includegraphics[width=0.8\textwidth]{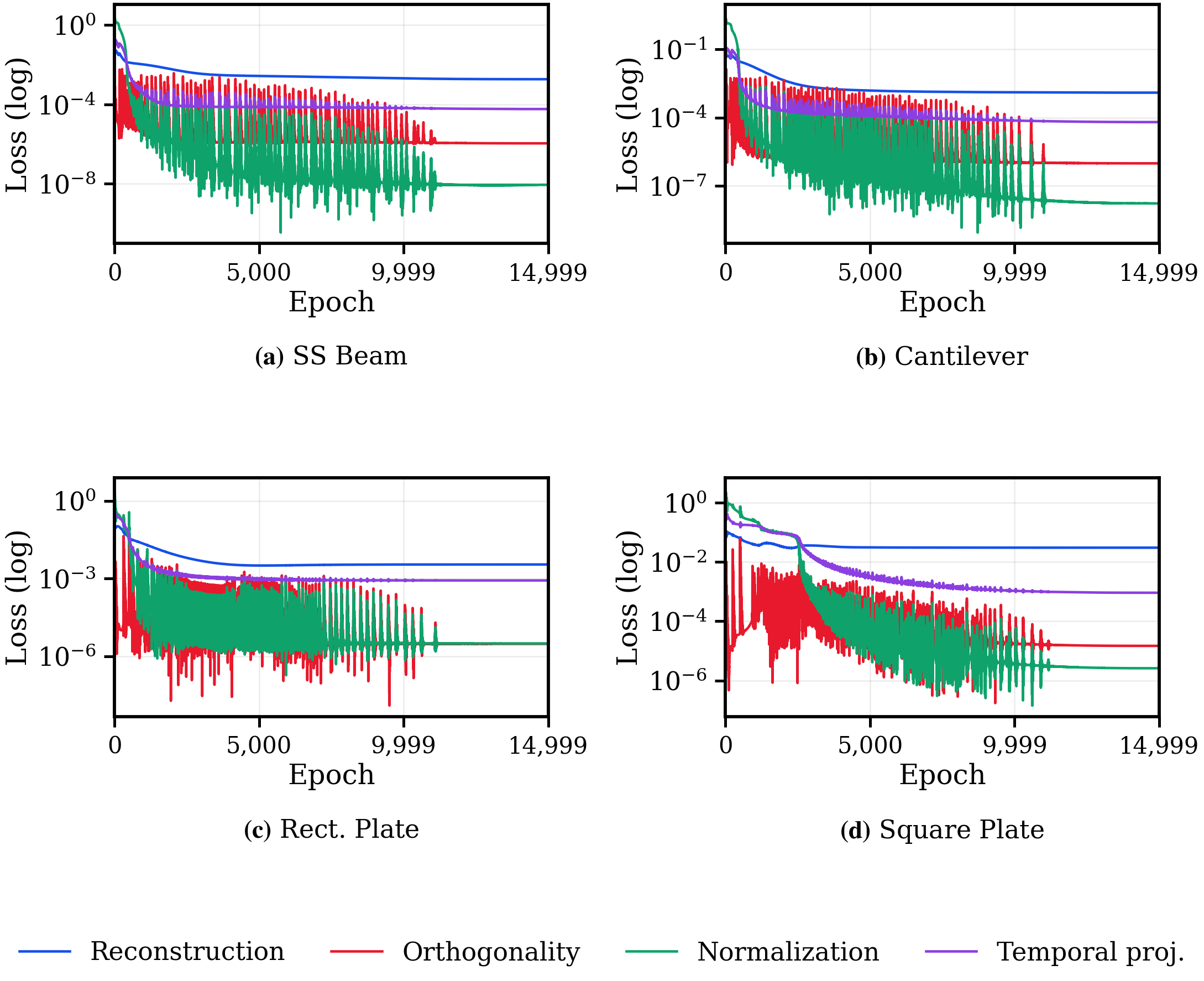}
    \caption{Training-loss convergence (log scale) for the four objective terms across the four structural systems: (a) simply supported Euler--Bernoulli beam, (b) cantilever beam, (c) rectangular Kirchhoff plate, and (d) square Kirchhoff plate; the reconstruction, orthogonality, normalization, and temporal projection losses all decrease steadily during training, with the temporal projection loss converging to the noise floor as the analytical modal basis is recovered.}
    \label{fig:loss}
  \vskip -0.15in
\end{figure*}

\Cref{fig:loss} shows the convergence of the four loss terms for each system on a log scale. The reconstruction and normalization terms fall steadily and then plateau. The orthogonality term is not monotone, and for a benign reason: at initialization the untrained trunk outputs are near zero, so every Gram entry---and hence $\mathcal{L}_{\mathrm{orth}}$---starts small; it rises sharply over the first few hundred iterations as the shapes grow toward unit norm, and is then driven back down. Its converged magnitude, rather than its trajectory, is what certifies the constraint, and the near-identity Gram matrices of \cref{fig:gram} confirm it. The temporal projection term \cref{eq:proj} falls to the noise floor as the modes settle, indicating that at convergence each learned shape is consistent with the field projected onto its own time signature. One fixed set of weights and a single schedule suffice for all four systems; no per-system tuning was required.

\begin{table}[t]
  \caption{Recovered modal basis. For each system and mode we report the analytical and learned natural frequency (rad/s) and the true and refined damping ratios, the latter obtained by the log-envelope regression of \cref{eq:modalproj}. The square-plate pair shares one tied frequency.}
  \label{tab:corr}
  \vspace{0.1in}
  \begin{small}
  \setlength{\tabcolsep}{8pt}
  \begin{tabular}{@{}llcccc@{}}
    \toprule
    System & Mode & $\omega$ & $\hat{\omega}$ & $\zeta$ & $\hat{\zeta}$ \\
    \midrule
    SS beam      & 1 & 9.87  & 9.87  & 0.020 & 0.0200 \\
                 & 2 & 39.5  & 39.1  & 0.030 & 0.0303 \\
                 & 3 & 88.8  & 87.1  & 0.025 & 0.0254 \\
    \midrule
    Cantilever   & 1 & 3.52  & 3.52  & 0.020 & 0.0200 \\
                 & 2 & 22.0  & 21.8  & 0.030 & 0.0303 \\
                 & 3 & 61.7  & 60.5  & 0.025 & 0.0255 \\
    \midrule
    Rect.\ plate & $(1,1)$ & 2.85 & 2.87 & 0.020 & 0.0201 \\
                 & $(2,1)$ & 5.48 & 5.40 & 0.030 & 0.0304 \\
                 & $(1,2)$ & 8.77 & 8.67 & 0.025 & 0.0253 \\
    \midrule
    Square plate & $(1,1)$ & 3.95 & 3.97  & 0.020 & 0.0192 \\
                 & $(2,1)$ & 9.87 & 10.35 & 0.030 & 0.0286 \\
                 & $(1,2)$ & 9.87 & 10.35 & 0.025 & 0.0234 \\
    \bottomrule
  \end{tabular}
  \end{small}
\end{table}

\subsection{Ablation Study}
\label{sec:ablation}

\begin{figure*}[tbp]
  \vskip 0.1in
  \centering
    \includegraphics[width=0.8\textwidth]{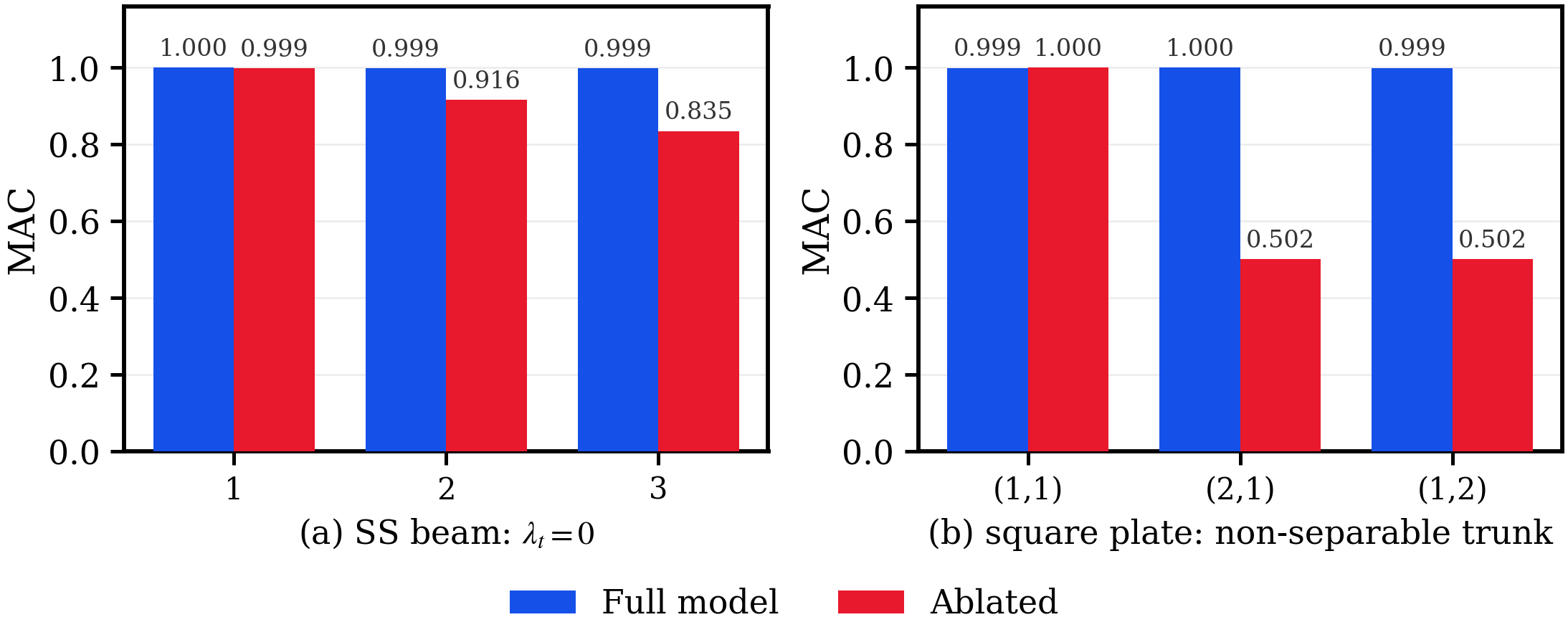}
    \caption{Ablation study comparing the full ModalONet model and ablated variants using the per-mode modal assurance criterion (MAC): (a) removing the temporal projection loss primarily degrades recovery of the fast-decaying beam modes, while the slowest mode remains largely unaffected. (b) replacing the separable trunk with a non-separable architecture causes the degenerate $(1,2)$ and $(2,1)$ square-plate modes to collapse into rotated mixtures (MAC $\approx 0.50 \approx \cos^2 45^\circ$), whereas the non-degenerate $(1,1)$ mode is recovered accurately.}
    \label{fig:ablation}
  \vskip -0.15in
\end{figure*}

Two components carry the method beyond a generic reconstructing DeepONet: the temporal projection term \cref{eq:proj} and the separable trunk \cref{eq:separable}. \Cref{fig:ablation} isolates each.

Removing the projection term ($\lambda_t{=}0$) leaves reconstruction, orthogonality, and normalization intact, yet the beam's faster-decaying modes degrade sharply: mode~3 falls from MAC $0.999$ to $0.835$ and mode~2 to $0.916$, while the slow mode~1 is unaffected (\cref{fig:ablation}a). This is the predicted failure. Without energy normalization, a briefly excited mode contributes little to the time-averaged reconstruction loss and drifts toward the dominant mode.

Replacing the separable trunk with a fully connected trunk of comparable capacity leaves the non-degenerate square-plate mode intact (MAC $1.000$) but collapses the degenerate pair to MAC $0.502$ each (\cref{fig:ablation}b)---almost exactly $\cos^{2}45^{\circ}$, the signature of a $45^{\circ}$ rotated mixture. The joint network may return any linear combination within the degenerate subspace while
still satisfying reconstruction and orthogonality, and this is exactly the case anticipated in \cref{sec:gram}: the learned Gram matrix remains near-identity while MAC collapses, so orthonormality alone cannot detect the failure. The tied frequency is unaffected ($4.8\%$), isolating the collapse to the spatial factorization. Each component thus addresses a distinct failure, and both are necessary.

\vspace{0.2cm}
\subsection{Interpretability of the recovered basis}
\label{sec:interpret}

Taken together, the preceding results establish that a neural operator can return a structural modal basis rather than a forward response. Across beams and plates, the method recovers mode shapes at MAC $\geq 0.998$ (\cref{fig:corr}), natural frequencies within $5\%$ and refined damping ratios within $7\%$ (\cref{tab:corr,fig:freq,fig:damprec}), on a learned basis that is orthonormal to within a few thousandths (\cref{fig:gram}).

Two properties distinguish this representation from prior learned modal identification. First, the mode shapes are continuous functions of the spatial coordinate rather than vectors tied to a sensor grid, so they can be evaluated at any resolution, including at points where no measurement was taken. Second, frequency and damping are native pole parameters obtained jointly with the shapes; the log-envelope regression that sharpens damping consumes only the learned shapes and the measured field, so no separate identification machinery is introduced at any stage.

The consequence is that every latent dimension of the operator is a physical mode, so the representation is interpretable and independently verifiable against theory---in contrast to forward operator-learning surrogates, whose latent factors may be accurate in aggregate yet carry no individual physical meaning. The architectural choices that deliver this are load-bearing rather than incidental, as \cref{sec:ablation} confirms: removing either the temporal projection term or the separable trunk degrades precisely the modes each was introduced to protect.

\subsection{Limitations}
\label{sec:limitations}

Among the three modal quantities considered, natural frequencies and damping ratios are more challenging to estimate than the mode shapes. For \emph{exactly} degenerate modes, the individual natural frequencies are not identifiable from response data. Instead, only the shared natural frequency and the associated two-dimensional modal subspace can be uniquely recovered. Accordingly, ModalONet assigns a single natural frequency to each degenerate pair, and the residual error reported in \cref{fig:freq} is computed with respect to this shared estimate. The adopted $2\%$ grouping tolerance is appropriate for symmetry-induced exact degeneracy but may require adjustment for closely spaced, non-degenerate modes. Damping ratios are also only weakly constrained by the reconstruction objective. Although the post-hoc log-envelope refinement substantially improves estimation accuracy, it assumes freely decaying responses with proportional damping and remains sensitive to errors in the estimated natural frequency, as illustrated by the square-plate example.

The present study is limited to synthetic data and a controlled problem setting. All experiments use responses generated by modal superposition under proportional damping with a known modal rank~$r$. Practical applications involving measured responses present additional challenges, including non-proportional damping, correlated measurement noise, and model-order selection, which are beyond the scope of the present study. Furthermore, ModalONet is trained independently for each structure. Consequently, each new system requires a separate optimization, and the current formulation exploits the DeepONet architecture primarily as a structured modal factorization rather than as a transferable operator across families of parameterized structures. Extending the framework to learn transferable representations across varying geometries, materials, and boundary conditions represents a promising direction for future research.



\section{Conclusion}
\label{sec:conclusion}

This work introduced ModalONet, a neural operator that recasts modal identification as an operator-learning problem, directly recovering the modal basis of a structure rather than its forward response. The formulation exploits the algebraic equivalence between the DeepONet branch--trunk decomposition and classical modal superposition, enabling the network to identify modal coordinates and mode shapes directly from sampled response fields. To ensure a physically meaningful decomposition, ModalONet combines reconstruction with orthonormality and temporal projection constraints, while a Laplace branch provides closed-form estimates of natural frequencies and damping ratios. The framework requires only measured response fields for training, without analytical mode shapes or conventional spectral and time--frequency preprocessing.

Across four benchmark structural systems, ModalONet recovered all twelve mode shapes with MAC values of at least $0.998$, preserved orthonormality to within $3\times10^{-3}$, estimated natural frequencies within $4.9\%$, and achieved damping-ratio errors below $6.5\%$ following log-envelope refinement. Beyond these quantitative results, two broader observations emerged. First, orthonormality alone is insufficient to guarantee correct modal recovery: a non-separable trunk preserved a nearly orthonormal basis while producing rotated mixtures of degenerate modes, highlighting the need to resolve eigenspace ambiguities explicitly. Second, the three modal quantities are not equally constrained by reconstruction. Whereas mode shapes and natural frequencies are accurately identified through the operator-learning formulation, reliable damping estimation benefits from a dedicated post-processing step.

The current formulation is trained independently for each structure, using the DeepONet factorization as a structured modal representation rather than as a transferable operator across structural families. Extending the framework to parameterized geometries, materials, and boundary conditions through conditioned neural operators represents a natural next step and would fully realize the advantages of operator learning. Future work will also focus on experimental validation using measured vibration data and on relaxing the assumptions discussed in \cref{sec:limitations}, enabling ModalONet to address practical modal identification problems under realistic operating conditions.


\begin{thebibliography}{39}
\providecommand{\natexlab}[1]{#1}
\providecommand{\url}[1]{\texttt{#1}}
\expandafter\ifx\csname urlstyle\endcsname\relax
  \providecommand{\doi}[1]{doi: #1}\else
  \providecommand{\doi}{doi: \begingroup \urlstyle{rm}\Url}\fi

\bibitem[Bao et~al.(2024)Bao, Liu, and Li]{bao2024mechanics}
Yuequan Bao, Dawei Liu, and Hui Li.
\newblock A mechanics-informed neural network method for structural modal
  identification.
\newblock \emph{Mechanical Systems and Signal Processing}, 216:\penalty0
  111458, 2024.
\newblock \doi{10.1016/j.ymssp.2024.111458}.

\bibitem[Ben-Shaul et~al.(2023)Ben-Shaul, Bar, Fishelov, and
  Sochen]{benshaul2023deep}
Ido Ben-Shaul, Leah Bar, Dalia Fishelov, and Nir Sochen.
\newblock Deep learning solution of the eigenvalue problem for differential
  operators.
\newblock \emph{Neural Computation}, 35\penalty0 (6):\penalty0 1100--1134,
  2023.
\newblock \doi{10.1162/neco_a_01583}.

\bibitem[Berkooz et~al.(1993)Berkooz, Holmes, and Lumley]{berkooz1993pod}
Gal Berkooz, Philip Holmes, and John~L. Lumley.
\newblock The proper orthogonal decomposition in the analysis of turbulent
  flows.
\newblock \emph{Annual Review of Fluid Mechanics}, 25\penalty0 (1):\penalty0
  539--575, 1993.
\newblock \doi{10.1146/annurev.fl.25.010193.002543}.

\bibitem[Brincker et~al.(2001)Brincker, Zhang, and Andersen]{brincker2001modal}
Rune Brincker, Lingmi Zhang, and Palle Andersen.
\newblock Modal identification of output-only systems using frequency domain
  decomposition.
\newblock \emph{Smart Materials and Structures}, 10\penalty0 (3):\penalty0
  441--445, 2001.
\newblock \doi{10.1088/0964-1726/10/3/303}.

\bibitem[Cao et~al.(2024{\natexlab{a}})Cao, Goswami, and
  Karniadakis]{cao2024laplace}
Qianying Cao, Somdatta Goswami, and George~Em Karniadakis.
\newblock Laplace neural operator for solving differential equations.
\newblock \emph{Nature Machine Intelligence}, 6\penalty0 (6):\penalty0
  631--640, 2024{\natexlab{a}}.
\newblock \doi{10.1038/s42256-024-00844-4}.

\bibitem[Cao et~al.(2024{\natexlab{b}})Cao, Goswami, Tripura, Chakraborty, and
  Karniadakis]{cao2024floating}
Qianying Cao, Somdatta Goswami, Tapas Tripura, Souvik Chakraborty, and
  George~Em Karniadakis.
\newblock Deep neural operators can predict the real-time response of floating
  offshore structures under irregular waves.
\newblock \emph{Computers \& Structures}, 291:\penalty0 107228,
  2024{\natexlab{b}}.
\newblock \doi{10.1016/j.compstruc.2023.107228}.

\bibitem[Chen and Chen(1995)]{chen1995universal}
Tianping Chen and Hong Chen.
\newblock Universal approximation to nonlinear operators by neural networks
  with arbitrary activation functions and its application to dynamical systems.
\newblock \emph{IEEE Transactions on Neural Networks}, 6\penalty0 (4):\penalty0
  911--917, 1995.
\newblock \doi{10.1109/72.392253}.

\bibitem[De and Brewick(2024)]{de2024bifidelity}
Subhayan De and Patrick~T. Brewick.
\newblock A bi-fidelity {DeepONet} approach for modeling hysteretic systems
  under uncertainty.
\newblock \emph{Applied Mathematical Modelling}, 135:\penalty0 708--728, 2024.
\newblock \doi{10.1016/j.apm.2024.06.045}.

\bibitem[Deng et~al.(2022)Deng, Shi, and Zhu]{deng2022neuralef}
Zhijie Deng, Jiaxin Shi, and Jun Zhu.
\newblock {NeuralEF}: Deconstructing kernels by deep neural networks.
\newblock In \emph{International Conference on Machine Learning (ICML)}, pages
  4976--4992. PMLR, 2022.

\bibitem[Ewins(2000)]{ewins2000modal}
David~J. Ewins.
\newblock \emph{Modal Testing: Theory, Practice and Application}.
\newblock Research Studies Press, 2nd edition, 2000.
\newblock ISBN 978-0863802188.

\bibitem[Feeny \& Kappagantu(1998)Feeny and Kappagantu]{feeny1998pod}
Feeny, B.~F. and Kappagantu, R.
\newblock On the physical interpretation of proper orthogonal modes in
  vibrations.
\newblock {\em Journal of Sound and Vibration}, 211(4):607--616, 1998.
\newblock \doi{10.1006/jsvi.1997.1386}.

\bibitem[Garg et~al.(2022)Garg, Gupta, and Chakraborty]{garg2022assessment}
Shailesh Garg, Harshit Gupta, and Souvik Chakraborty.
\newblock Assessment of {DeepONet} for time-dependent reliability analysis of
  dynamical systems subjected to stochastic loading.
\newblock \emph{Engineering Structures}, 270:\penalty0 114811, 2022.
\newblock \doi{10.1016/j.engstruct.2022.114811}.

\bibitem[Giurgiutiu(2014)]{giurgiutiu2014shm}
Victor Giurgiutiu.
\newblock \emph{Structural Health Monitoring with Piezoelectric Wafer Active
  Sensors}.
\newblock Academic Press, Elsevier, Oxford, UK, 2nd edition, 2014.
\newblock ISBN 978-0-12-418691-0.

\bibitem[Goswami et~al.(2022)Goswami, Yin, Yu, and
  Karniadakis]{goswami2022variational}
Somdatta Goswami, Minglang Yin, Yue Yu, and George~Em Karniadakis.
\newblock A physics-informed variational {DeepONet} for predicting crack path
  in quasi-brittle materials.
\newblock \emph{Computer Methods in Applied Mechanics and Engineering},
  391:\penalty0 114587, 2022.
\newblock \doi{10.1016/j.cma.2022.114587}.

\bibitem[Goswami et~al.(2025)Goswami, Giovanis, Li, Spence, and
  Shields]{goswami2025neural}
Somdatta Goswami, Dimitris~G. Giovanis, Bowei Li, Seymour M.~J. Spence, and
  Michael~D. Shields.
\newblock Neural operators for stochastic modeling of nonlinear structural
  system response to natural hazards.
\newblock \emph{Engineering Structures}, 345:\penalty0 121284, 2025.
\newblock \doi{10.1016/j.engstruct.2025.121284}.

\bibitem[Han et~al.(2020)Han, Lu, and Zhou]{han2020solving}
Jiequn Han, Jianfeng Lu, and Mo~Zhou.
\newblock Solving high-dimensional eigenvalue problems using deep neural
  networks: A diffusion {Monte Carlo} like approach.
\newblock \emph{Journal of Computational Physics}, 423:\penalty0 109792, 2020.
\newblock \doi{10.1016/j.jcp.2020.109792}.

\bibitem[He et~al.(2024)He, Kushwaha, Park, Koric, Abueidda, and
  Jasiuk]{he2024sequential}
Junyan He, Shashank Kushwaha, Jaewan Park, Seid Koric, Diab Abueidda, and Iwona
  Jasiuk.
\newblock Sequential deep operator networks ({S-DeepONet}) for predicting
  full-field solutions under time-dependent loads.
\newblock \emph{Engineering Applications of Artificial Intelligence},
  127:\penalty0 107258, 2024.
\newblock \doi{10.1016/j.engappai.2023.107258}.

\bibitem[Hern{\'a}ndez-Gonz{\'a}lez et~al.(2024)Hern{\'a}ndez-Gonz{\'a}lez,
  Garc{\'i}a-Mac{\'i}as, Costante, and Ubertini]{hernandez2024blind}
Iv{\'a}n~A. Hern{\'a}ndez-Gonz{\'a}lez, Enrique Garc{\'i}a-Mac{\'i}as, Gabriele
  Costante, and Filippo Ubertini.
\newblock {AI}-driven blind source separation for fast operational modal
  analysis of structures.
\newblock \emph{Mechanical Systems and Signal Processing}, 211:\penalty0
  111267, 2024.
\newblock \doi{10.1016/j.ymssp.2024.111267}.

\bibitem[Jian et~al.(2025{\natexlab{a}})Jian, Bacsa, Duth{\'e}, and
  Chatzi]{jian2025modal}
Xudong Jian, Kiran Bacsa, Gregory Duth{\'e}, and Eleni Chatzi.
\newblock Modal decomposition and identification for a population of structures
  using physics-informed graph neural networks and transformers.
\newblock \emph{Mechanical Systems and Signal Processing}, 241:\penalty0
  113604, 2025{\natexlab{a}}.
\newblock \doi{10.1016/j.ymssp.2025.113604}.

\bibitem[Jian et~al.(2025{\natexlab{b}})Jian, Xia, Duth{\'e}, Bacsa, Liu, and
  Chatzi]{jian2025gnn}
Xudong Jian, Yutong Xia, Gregory Duth{\'e}, Kiran Bacsa, Wei Liu, and Eleni
  Chatzi.
\newblock Using graph neural networks and frequency domain data for automated
  operational modal analysis of populations of structures.
\newblock \emph{Data-Centric Engineering}, 6:\penalty0 e45, 2025{\natexlab{b}}.
\newblock \doi{10.1017/dce.2025.10023}.

\bibitem[Juang \& Pappa(1985)Juang and Pappa]{juang1985era}
Juang, J.-N. and Pappa, R.~S.
\newblock An eigensystem realization algorithm for modal parameter
  identification and model reduction.
\newblock {\em Journal of Guidance, Control, and Dynamics}, 8(5):620--627,
  1985.
\newblock \doi{10.2514/3.20031}.

\bibitem[Kaewnuratchadasorn et~al.(2024)Kaewnuratchadasorn, Wang, and
  Kim]{kaewnuratchadasorn2024neural}
Chawit Kaewnuratchadasorn, Jiaji Wang, and Chul-Woo Kim.
\newblock Neural operator for structural simulation and bridge health
  monitoring.
\newblock \emph{Computer-Aided Civil and Infrastructure Engineering},
  39\penalty0 (6):\penalty0 872--890, 2024.
\newblock \doi{10.1111/mice.13105}.

\bibitem[Kovachki et~al.(2023)Kovachki, Li, Liu, Azizzadenesheli, Bhattacharya,
  Stuart, and Anandkumar]{kovachki2023neural}
Nikola Kovachki, Zongyi Li, Burigede Liu, Kamyar Azizzadenesheli, Kaushik
  Bhattacharya, Andrew Stuart, and Anima Anandkumar.
\newblock Neural operator: Learning maps between function spaces with
  applications to {PDEs}.
\newblock \emph{Journal of Machine Learning Research}, 24\penalty0
  (89):\penalty0 1--97, 2023.

\bibitem[Lai et~al.(2022)Lai, Liu, Jian, Bacsa, Sun, and Chatzi]{lai2022neural}
Zhilu Lai, Wei Liu, Xudong Jian, Kiran Bacsa, Limin Sun, and Eleni Chatzi.
\newblock Neural modal ordinary differential equations: Integrating
  physics-based modeling with neural ordinary differential equations for
  modeling high-dimensional monitored structures.
\newblock \emph{Data-Centric Engineering}, 3:\penalty0 e34, 2022.
\newblock \doi{10.1017/dce.2022.35}.

\bibitem[Li et~al.(2021)Li, Kovachki, Azizzadenesheli, Liu, Bhattacharya,
  Stuart, and Anandkumar]{li2021fourier}
Zongyi Li, Nikola Kovachki, Kamyar Azizzadenesheli, Burigede Liu, Kaushik
  Bhattacharya, Andrew Stuart, and Anima Anandkumar.
\newblock Fourier neural operator for parametric partial differential
  equations.
\newblock In \emph{International Conference on Learning Representations
  (ICLR)}, 2021.
\newblock \doi{10.48550/arXiv.2010.08895}.

\bibitem[Liu and Bao(2025)]{liu2025closely}
Dawei Liu and Yuequan Bao.
\newblock A mechanics-informed neural network method for structural modal
  identification: Application to closely spaced modes.
\newblock \emph{Journal of Sound and Vibration}, 612:\penalty0 119154, 2025.
\newblock \doi{10.1016/j.jsv.2025.119154}.

\bibitem[Liu et~al.(2021)Liu, Tang, Bao, and Li]{liu2021blind}
Dawei Liu, Zhiyi Tang, Yuequan Bao, and Hui Li.
\newblock Machine-learning-based methods for output-only structural modal
  identification.
\newblock \emph{Structural Control and Health Monitoring}, 28\penalty0
  (12):\penalty0 e2843, 2021.
\newblock \doi{10.1002/stc.2843}.

\bibitem[Liu et~al.(2024{\natexlab{a}})Liu, Nath, and Cai]{liu2024causality}
Lizuo Liu, Kamaljyoti Nath, and Wei Cai.
\newblock A causality-{DeepONet} for causal responses of linear dynamical
  systems.
\newblock \emph{Communications in Computational Physics}, 35\penalty0
  (5):\penalty0 1194--1228, 2024{\natexlab{a}}.
\newblock \doi{10.4208/cicp.OA-2023-0078}.

\bibitem[Liu et~al.(2024{\natexlab{b}})Liu, Cao, Xu, and Deng]{liu2024cvmodal}
Yingkai Liu, Ran Cao, Shaopeng Xu, and Lu~Deng.
\newblock A deep learning-based method for structural modal analysis using
  computer vision.
\newblock \emph{Engineering Structures}, 301:\penalty0 117285,
  2024{\natexlab{b}}.
\newblock \doi{10.1016/j.engstruct.2023.117285}.

\bibitem[Lu et~al.(2021)Lu, Jin, Pang, Zhang, and Karniadakis]{lu2021learning}
Lu~Lu, Pengzhan Jin, Guofei Pang, Zhongqiang Zhang, and George~Em Karniadakis.
\newblock Learning nonlinear operators via {DeepONet} based on the universal
  approximation theorem of operators.
\newblock \emph{Nature Machine Intelligence}, 3\penalty0 (3):\penalty0
  218--229, 2021.
\newblock \doi{10.1038/s42256-021-00302-5}.

\bibitem[Lu et~al.(2022)Lu, Meng, Cai, Mao, Goswami, Zhang, and
  Karniadakis]{lu2022comprehensive}
Lu~Lu, Xuhui Meng, Shengze Cai, Zhiping Mao, Somdatta Goswami, Zhongqiang
  Zhang, and George~Em Karniadakis.
\newblock A comprehensive and fair comparison of two neural operators (with
  practical extensions) based on {FAIR} data.
\newblock \emph{Computer Methods in Applied Mechanics and Engineering},
  393:\penalty0 114778, 2022.
\newblock \doi{10.1016/j.cma.2022.114778}.

\bibitem[Mandl et~al.(2025)Mandl, Goswami, Lambers, and
  Ricken]{mandl2024separable}
Luis Mandl, Somdatta Goswami, Lena Lambers, and Tim Ricken.
\newblock Separable physics-informed {DeepONet}: Breaking the curse of
  dimensionality in physics-informed machine learning.
\newblock \emph{Computer Methods in Applied Mechanics and Engineering},
  434:\penalty0 117586, 2025.
\newblock \doi{10.1016/j.cma.2024.117586}.

\bibitem[Peeters and De~Roeck(2001)]{peeters2001stochastic}
Bart Peeters and Guido De~Roeck.
\newblock Stochastic system identification for operational modal analysis: A
  review.
\newblock \emph{Journal of Dynamic Systems, Measurement, and Control},
  123\penalty0 (4):\penalty0 659--667, 2001.
\newblock \doi{10.1115/1.1410370}.

\bibitem[Raissi et~al.(2019)Raissi, Perdikaris, and
  Karniadakis]{raissi2019physics}
Maziar Raissi, Paris Perdikaris, and George~Em Karniadakis.
\newblock Physics-informed neural networks: A deep learning framework for
  solving forward and inverse problems involving nonlinear partial differential
  equations.
\newblock \emph{Journal of Computational Physics}, 378:\penalty0 686--707,
  2019.
\newblock \doi{10.1016/j.jcp.2018.10.045}.

\bibitem[Sadhu et~al.(2017)Sadhu, Narasimhan, and Antoni]{sadhu2017review}
Ayan Sadhu, Sriram Narasimhan, and J{\'e}r{\^o}me Antoni.
\newblock A review of output-only structural mode identification literature
  employing blind source separation methods.
\newblock \emph{Mechanical Systems and Signal Processing}, 94:\penalty0
  415--431, 2017.
\newblock \doi{10.1016/j.ymssp.2017.03.001}.

\bibitem[Schmid(2010)]{schmid2010dmd}
Peter~J. Schmid.
\newblock Dynamic mode decomposition of numerical and experimental data.
\newblock \emph{Journal of Fluid Mechanics}, 656:\penalty0 5--28, 2010.
\newblock \doi{10.1017/S0022112010001217}.

\bibitem[Su et~al.(2020)Su, Huang, Song, and LaFave]{su2020automatic}
Liang Su, Xin Huang, Ming-liang Song, and James~Michael LaFave.
\newblock Automatic identification of modal parameters for structures based on
  an uncertainty diagram and a convolutional neural network.
\newblock \emph{Structures}, 28:\penalty0 369--379, 2020.
\newblock \doi{10.1016/j.istruc.2020.08.077}.

\bibitem[Wang et~al.(2021)Wang, Wang, and Perdikaris]{wang2021learning}
Sifan Wang, Hanwen Wang, and Paris Perdikaris.
\newblock Learning the solution operator of parametric partial differential
  equations with physics-informed {DeepONets}.
\newblock \emph{Science Advances}, 7\penalty0 (40):\penalty0 eabi8605, 2021.
\newblock \doi{10.1126/sciadv.abi8605}.

\bibitem[Yang et~al.(2026)Yang, Du, and Liu]{yang2026neural}
Zherui Yang, Tao Du, and Ligang Liu.
\newblock Learning {Laplacian} eigenspace with mass-aware neural operators on
  point clouds.
\newblock \emph{ACM Transactions on Graphics (Proc.\ SIGGRAPH)}, 2026.
\newblock \doi{10.48550/arXiv.2605.24390}.

\bibitem[Ahmed and Kopsaftopoulos(2025)]{ahmed2025functional}
Shabbir Ahmed and Fotis Kopsaftopoulos.
\newblock A functional series time-dependent framework for non-stationary modeling and statistical damage diagnosis via ultrasonic guided waves.
\newblock \emph{Structural Health Monitoring}, \penalty0 14759217251369342, 2025.

\bibitem[Ahmed et~al.(2025)Ahmed, Farhangdoust, and Chang]{ahmed2025autoregressive}
Shabbir Ahmed, Saman Farhangdoust, and Fu-Kuo Chang.
\newblock Autoregressive model-based parameter correlation for state of charge and state of health of lithium-ion batteries using built-in piezoelectric transducer induced ultrasonic waves.
\newblock \emph{Journal of Energy Storage}, 114:\penalty0 115829, 2025.

\bibitem[Ahmed and Kopsaftopoulos(2024)]{ahmed2024active}
Shabbir Ahmed and Fotis Kopsaftopoulos.
\newblock Active sensing ultrasonic guided wave-based damage diagnosis via stochastic stationary time-series models.
\newblock \emph{Structural Health Monitoring}, 23\penalty0 (4):2559--2595, 2024.

\bibitem[Rizmi and Ahmed(2026)]{rizmi2026vision}
R.~K.~B.~M. Rizmi and Shabbir Ahmed.
\newblock Vision-based structural damage identification in vibrating beams via dynamic mode decomposition.
\newblock \emph{arXiv preprint}, arXiv:\penalty0 2605.02803, 2026.

\end{thebibliography}
\end{document}